\documentclass[letter]{aa} 
\usepackage{subcaption}
\usepackage{graphicx}
\usepackage{txfonts}
\usepackage{color}
\usepackage{float}
\usepackage{soul}
\begin{document} 

   \title{An unbiased NOEMA 2.6 to 4\,mm survey of the GG Tau ring: \\
   First detection of CCS in a protoplanetary disk}
%
%
   \author {N.T. Phuong \inst{1,2}
    \and A. Dutrey \inst{3}
    \and  E. Chapillon \inst{3,4}
    \and S. Guilloteau \inst{3} 
    \and J. Bary \inst{5}
    \and T. L. Beck \inst{6} 
    \and A. Coutens \inst{7} 
    \and O. Denis-Alpizar \inst{8} 
    \and E. Di Folco \inst{3} 
    \and P.N. Diep \inst{2} 
    \and L. Majumdar \inst{9} 
    \and J-P. Melisse \inst{3,4} 
    \and C-W. Lee \inst{1,10}
    \and V. Pietu \inst{4} 
    \and T. Stoecklin \inst{11}
    \and Y-W. Tang \inst{12} 
    }
 \institute{Korea Astronomy and  Space Science Institute, 776 Daedeokdae-ro, Yuseong-gu, Daejeon, Korea;$^\star$tpnguyen@kasi.re.kr 
                 \and Department of Astrophysics, Vietnam National Space Center, 
                        Vietnam Academy of Science and Techonology, 18 Hoang Quoc Viet, Cau Giay, Hanoi, Vietnam 
                \and  Laboratoire d'Astrophysique de Bordeaux, Universit\'e de Bordeaux, CNRS, B18N, All\'ee Geoffroy Saint-Hilaire, F-33615 Pessac 
                 \and IRAM, 300 rue de la piscine, F-38406 Saint Martin d'H\`eres Cedex, France
                 \and Department of Physics and Astronomy, Colgate University, 13 Oak Drive, Hamilton, New York 13346, USA
                 \and Space Telescope Science Institute, 3700 San Martin Drive, Baltimore, Maryland 21218, USA   
                 \and Institut de Recherche en Astrophysique et Plan\'etologie, Universit\'e de Toulouse, UPS-OMP, CNRS, CNES, 9 av. du Colonel Roche, 31028 Toulouse Cedex 4, France
                 \and Instituto de Ciencias Qu\'imicas Aplicadas, Facultad de Ingenier\'ia, Universidad Aut\'onoma de Chile, Av. Pedro de Valdivia 425, 7500912 Providencia, Santiago, Chile
                 \and School of Earth and Planetary Sciences, National Institute of Science Education and Research, HBNI, Jatni 752050, Odisha, India
                 \and University of Science and Technology, 217 Gajeong-ro, Yuseong-gu, Daejeon 34113, Republic of Korea
                 \and Institut des Sciences Mol\'eculaires, UMR5255-CNRS, 351 Cours de la libration, F-33405 Talence France
                 \and Academia Sinica Institute of Astronomy and Astrophysics, PO Box 23-141, Taipei 106, Taiwan
}
 \date{Received ; accepted }

  \abstract
   {Molecular line surveys are among the main tools to probe the structure and physical conditions in protoplanetary disks (PPDs), the birthplace of planets. The large radial and vertical temperature as well as density gradients in these PPDs lead to a complex chemical composition, making chemistry an important step to understand the variety of planetary systems.}
   {We aimed to study the chemical content of the protoplanetary disk surrounding GG Tau A, a well-known triple T\,Tauri system.}
   {We used NOEMA with the new correlator PolyFix to observe rotational lines at $\sim$2.6 to 4\,mm from a few dozen molecules. We  analysed the data with a radiative transfer code to derive molecular densities and the abundance relative to $^{13}$CO, which we compare to those of the TMC1 cloud and LkCa\,15 disk.} 
   {We report the first detection of CCS in PPDs. We also marginally detect OCS and find 16 other molecules in the GG Tauri outer disk. Ten of them had been found previously, while seven others ($^{13}$CN, N$_2$H$^+$, HNC, DNC, HC$_3$N, CCS, and C$^{34}$S) are new detections in this disk.} 
   {The analysis confirms that sulphur chemistry is not yet properly understood. The D/H ratio, derived from DCO$^{+}$/HCO$^{+}$, DCN/HCN, and DNC/HNC ratios, points towards a low temperature chemistry. The detection of the rare species CCS confirms that GG Tau 
   is a good laboratory to study the protoplanetary disk chemistry, thanks to its large disk size and mass.}

   \keywords{Planetary systems: protoplanetary disks - Molecular data - Astrochemistry - Stars: individual (GG Tau A).}
   \titlerunning{An unbiased NOEMA 2.6 to 4 mm survey of the GG Tau ring}
    \maketitle
   
%
\section{Introduction}
\vspace{-0.25cm}
The chemical content in a protoplanetary disk (PPD) is thought to be a combination of parent cloud inheritances and the product of in situ reactions.
PPDs are flared and layered, displaying important radial and vertical temperature and density gradients that result in a complex chemical structure and evolution.  Each layer in the disk has conditions suitable for different chemical reactions, leading to different molecular abundances.  For example, photo sensitive molecules such as CN and CCH are believed to probe the upper most layer which is directly irradiated by stellar UV; CO and its isotopologues arise from the layer just below it, while molecules are frozen on the dust grains (millimetre-sized) settled in the cold disk midplane. So far, more than thirty molecules have been detected in PPDs (see \citet{Phuong+Chapillon+Majumdar_2018} for a list, and the more recent detections of H$_2$CS by \citet{LeGal+Oberg_2019}, DNC by \citet{Loomis+Oberg+Andrews_2020}, SO by \citep{Riviere-Marichalar+etal_2020}, and SO$_2$ by \citet{Booth+Marel+Leemker_2021}).

In this paper, we report the first deep survey of a PPD covering the 2.6 to 4.2\,mm window, where fundamental
transitions of most simple molecules occur. We observed the GG Tau A system with the NOEMA
interferometer.  GG Tau A is a triple T Tauri system (Aa-Ab1/b2) with respective separations of 35 and 4.5 au \citep{DiFolco+Dutrey+LeBouquin_2014}, located in the Taurus-Auriga star forming region (150\,pc, Gaia 2018). It is surrounded by a large and massive Keplerian circumbinary (or ternary) disk with an estimated
mass \mbox{($M_{\mathrm{disk}}$=0.15\,M$_\odot$)} using dust properties that are typically assumed
for PPDs.  The disk consists of a dense, narrow (radius from \mbox{$\sim$180 to 260\,au}) ring
that contains 70\% of the disk mass. 
Beyond the ring, the outer gas disk extends out to $\sim$800\,au. The temperature profile of the disk has been studied by \citet{Dutrey+DiFolco+Guilloteau_2014}, \citet{Guilloteau+Dutrey+Simon_1999}, and \citet{Phuong+Dutrey+Diep_2020}: the dust temperature is
14\,K at 200\,au, and the kinetic temperature derived from CO analysis is $\sim$25\,K at the same radius. The large size, low temperature, and large mass make GG Tau A disk an ideal laboratory to study cold molecular chemistry. 

We present the observations and results in Sec.\ref{sec:obs-results}. Data analysis using the radiative transfer code DiskFit is presented in Sec.\ref{sec:data-ana}. We then discuss the results in Sec.\ref{sec:dis}. 
\vspace{-0.5cm}
\section{Observations and results} 
\label{sec:obs-results}
\subsection{Observations}

Observations were carried out using the NOEMA array between Sep 2019 and Apr 2020 (Projects
S19AZ and W19AV). Four frequency setups were used to cover the frequency range \mbox{70.5--113.5\,GHz}. The PolyFix correlator covered the full bandwidth of $\sim$15\,GHz at 2\,MHz resolution (5--6.5\,km\,s$^{-1}$, while the total
line width of GG Tau A is around 3.5\,km\,s$^{-1}$, see Fig.\ref{fig:spec}), plus a large number
of spectral windows with a 62.5\,kHz channel spacing (resolution 0.2--0.3\,km\,s$^{-1}$, comparable
to the local thermal linewidth in GG Tau), optimised to cover a maximum number of spectral lines. 
The phase calibrators used were J0440+146 and 0507+179. Flux calibration was done using MWC349 and LkHa101.
The standard pipeline in the CLIC package was used for calibration, with minor data editing.
Three setups covering [70.5--78; 86--93.5], [80.5--88; 96--103.5], and \mbox{[90.5--98; 106.3--113.5]\,GHz}
used the C and D configuration, while the band [88.8--96.5; 104.3--112] used the C and A configuration,
providing enhanced sensitivity and angular resolution in the overlapping frequencies.
 In total, we covered at high spectral resolution of 70 rotational lines
of 38 molecular species.

\subsection{Data reduction and imaging}

\begin{figure}[htbp!]
 \centering  
 \includegraphics[width=\linewidth]{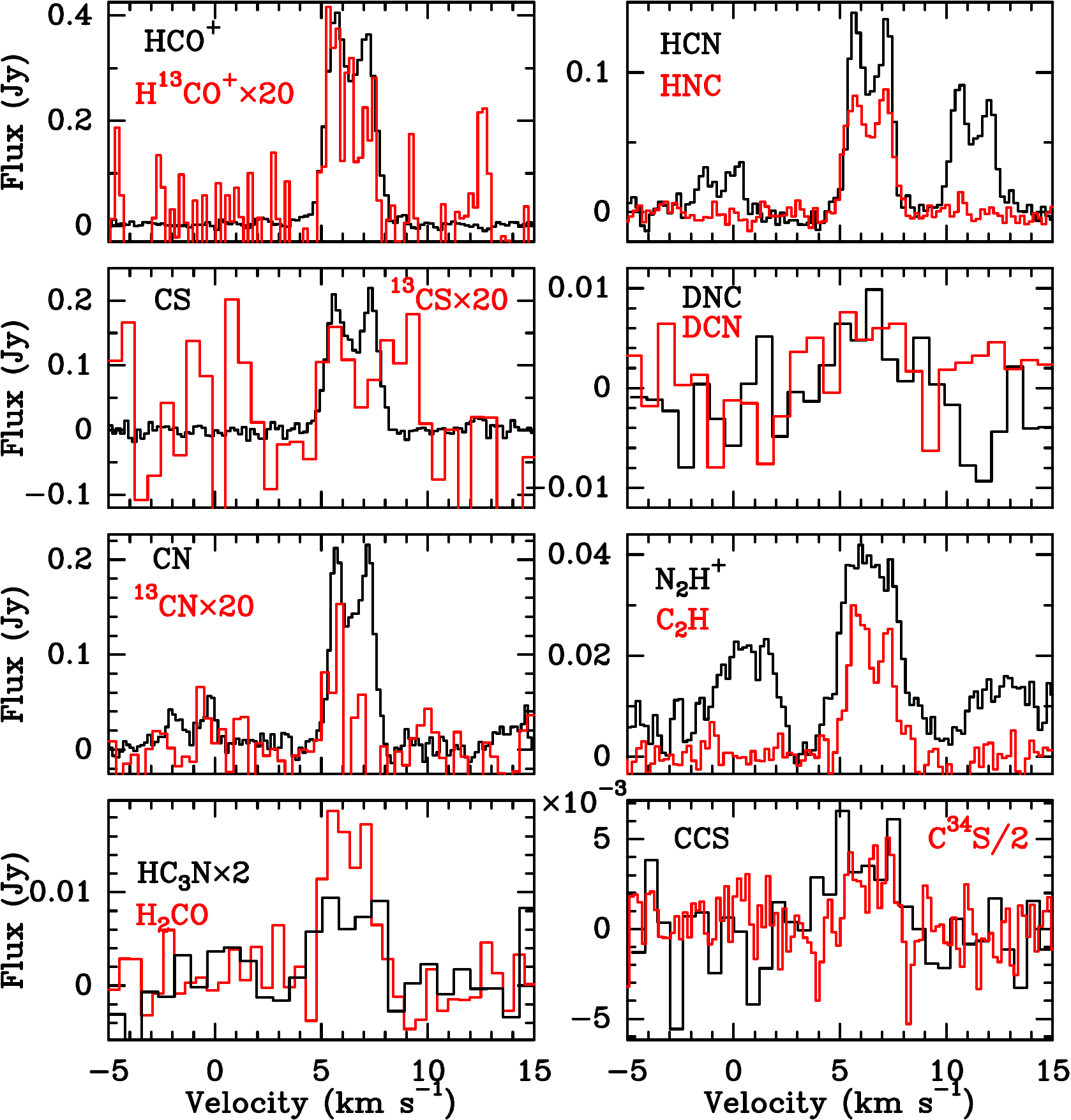}
    \caption{Spectrum of each detected molecule, integrated in the area of $15''\times15''$ for lines with S/N > 5 and in the 
    centre (central beam) of the detection area for lines with S/N<5. Some lines have been scaled up by a factor ($\times x)$ to give a better view.} \label{fig:spec}
\end{figure}

We used the IMAGER\footnote{https://imager.oasu.u-bordeaux.fr/} interferometric imaging package
to produce images. Continuum removal was carried out by the subtraction
of a zero-order frequency baseline for each visibility in the $uv$ plane.
Various combinations of natural or robust weightings were used in the image processing to highlight the image properties. Beam sizes and sensitivities are summarised in Table \ref{tab:beam} for all of the detected molecules. 
We note that HC$_3$N and CCS have faint lines coming from similar rotational levels and similar line strengths: to illustrate their detection, lines from each molecule were stacked in velocity to produce a single data set.  As synthesised beams exhibit significant sidelobes, deconvolution was done using the Hogbom method down to the rms noise.  Integrated spectra are shown in Fig.\ref{fig:spec}, and moment 0 maps are shown in Figs.\ref{fig:highSNR}-\ref{fig:mediumSNR}.
To further enhance the S/N, we used the Keplerian deprojection technique using kinematics and geometrical parameters from  \citet{Dutrey+DiFolco+Guilloteau_2014}, see Table \ref{tab:disk}. The images are deprojected and the velocities were corrected from the rotation pattern, thereby bringing all signals to the systemic velocity \citep[see][]{Teague+etal_2016,Yen+etal_2016}.
The resulting spectra and radius-velocity (RV) diagrams are shown in Fig.\ref{fig:lowSNR} and Fig.\ref{fig:rv}. The `teardrop' pattern  \citep{Teague_2019} expected for Keplerian disks transformed into `Eiffel tower'-like plots, because of the lack of emission from the tidal cavity of GG Tau A. Radial profiles of the line peak brightness are displayed in Appendix \ref{sec:prof}.

\begin{table}[htbp!]
\tiny
\caption{Obtained beam and noise of detected molecules}\label{tab:beam}
\centering 

\setlength\tabcolsep{1.5pt}
\begin{tabular}{lcccccc} 
\hline 
Molecules & Transition & Frequency & E$_u$/k  & Beam (PA) & Noise    & $\Delta V$ \\
                 &                &  (GHz)  &  (K)             & & (mJy/b) &  (km\,s$^{-1}$) \\
\hline\hline
CN & N=1-- 0 & 113.144 & 5.4 &$3.4''\times2.8'' (38^\circ)$ & 4.1 & 0.17 \\
HCN & J=1-- 0 & 88.632 & 4.3 &$3.5''\times2.3'' (18^\circ)$ & 1.7 &0.21 \\
HNC & J=1-- 0 & 90.664 & 4.4 & $3.2''\times2.0'' (18^\circ)$ & 1.4 &0.21\\
HCO$^+$ & J=1-- 0 & 89.189 & 4.3 & $2.0''\times1.2'' (22^\circ$) & 1.8 & 0.21\\
CS & J=2-- 1 & 97.981 & 7.1 &$4.2''\times3.2'' (1^\circ)$ & 4.0 &0.19\\ 
\hline
N$_2$H$^+$ &J=1--0  & 93.171 & 4.5 &$3.0''\times2.5'' (16^\circ)$ & 1.0 & 0.20 \\
CCH & N=1--0 & 87.317 & 4.2 & $3.0''\times1.8'' (19^\circ)$ & 0.6 &0.21 \\
H$^{13}$CO$^+$ & J=1-- 0 & 86.754 & 4.2 &$3.8''\times2.5'' (16^\circ)$ & 1.6 & 0.21 \\
$p$-H$_2$CO & $1_{( 0, 1)} - 0_{( 0, 0)}$ & 72.838   & 3.5 & $4.4''\times 2.8'' (18^\circ)$ & 2.0 & 0.25 \\
$o$-H$_2$CO & $6_{( 1, 5)} - 6_{( 1, 6)}$ & 101.333 & 57.5 & $5.0''\times 3.6'' (0^\circ)$ & 4.0 & 0.18 \\ 
                                          \hline
                                           & J= 8-- 7 & 72.784 & 15.7&  & &\\                                         
				         & J= 9-- 8 & 81.881 &19.6 & & & \\ 
                   HC$_3$N       & J=10-- 9 & 90.979 &24.0 & $3.5''\times2.2'' (17^\circ$) & 0.6 & 0.77 \\ 
                                          & J=11--10 & 100.076 &28.8 & & & \\ 
                                          & J=12--11& 109.172 &34.1 & & & \\ 
\hline
DCN &J=1-- 0 & 72.414 & 3.5 &$4.3''\times2.9'' (18^\circ)$& 1.5 &0.78\\ 
DNC & J=1-- 0 & 76.306 & 3.7  & $4.3''\times2.9'' (18^\circ)$ & 1.2 &0.74\\ 
$^{13}$CS & J=2-- 1 & 92.494 & 6.7& $3.0''\times1.8'' (18^\circ)$ & 0.8 &0.61\\
C$^{34}$S & J=2-- 1 & 96.413 & 6.7& $4.9''\times4.0'' (0^\circ)$ & 2.5 &0.20\\

$^{13}$CN & N=1--0 & 108.056 & 5.2& $2.3''\times1.3'' (15^\circ)$ & 2.0 &0.34\\
\hline
                                  & (6-- 5, 7-- 6) & 81.505  &15.4 & &  &\\ 
                         CCS & (7-- 6, 8-- 7) & 93.870 &19.9 & $3.5''\times1.9'' (14^\circ)$ & 1.0 & 0.60 \\ 
                                  & (8-- 7, 7-- 6) & 99.866 &28.1 & & & \\ 
                                  \hline

\hline 
\end{tabular}
\end{table} 

 \subsection{Results}
\begin{figure*}[h]
 \centering

\includegraphics[width=0.23\linewidth]{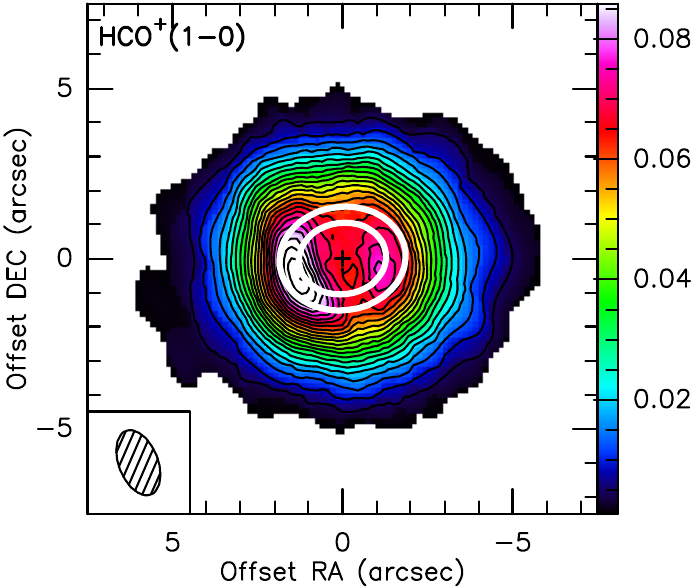}
\includegraphics[width=0.22\linewidth]{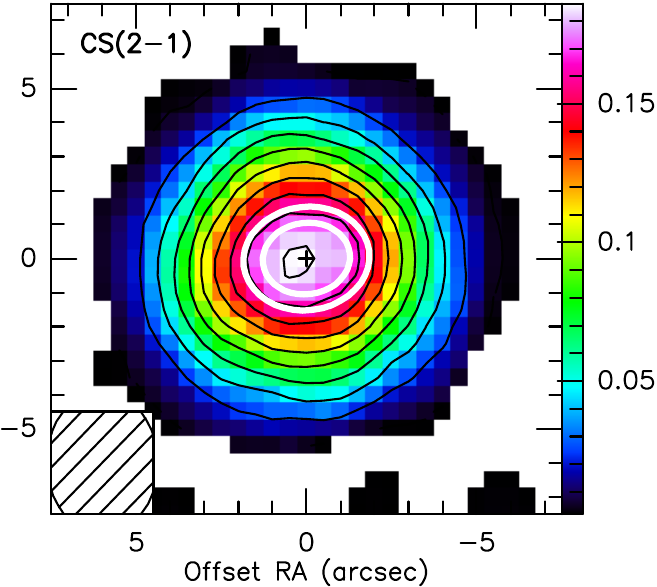}
\includegraphics[width=0.22\linewidth]{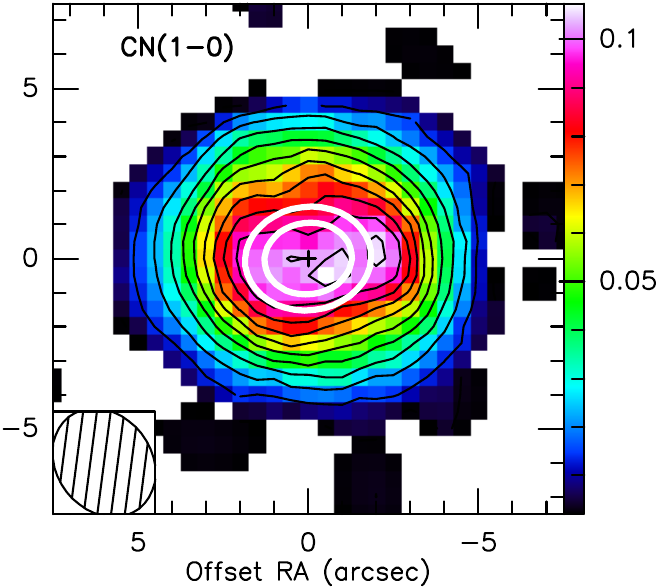}
\includegraphics[width=0.22\linewidth]{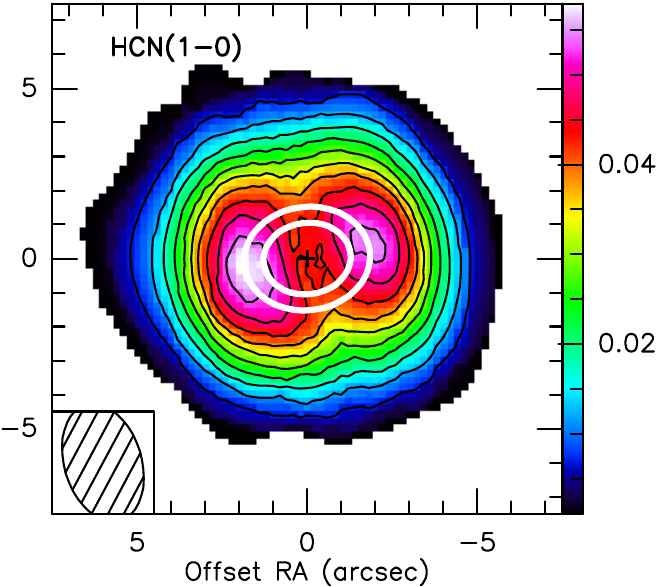}
\includegraphics[width=0.23\linewidth]{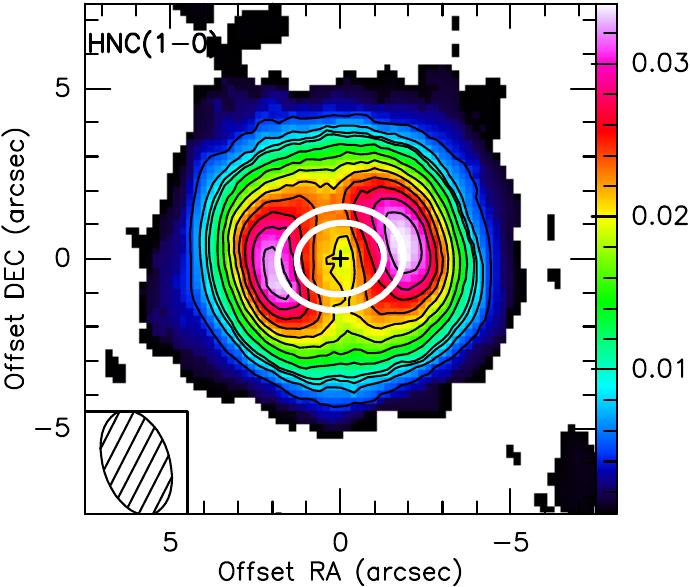}
\includegraphics[width=0.22\linewidth]{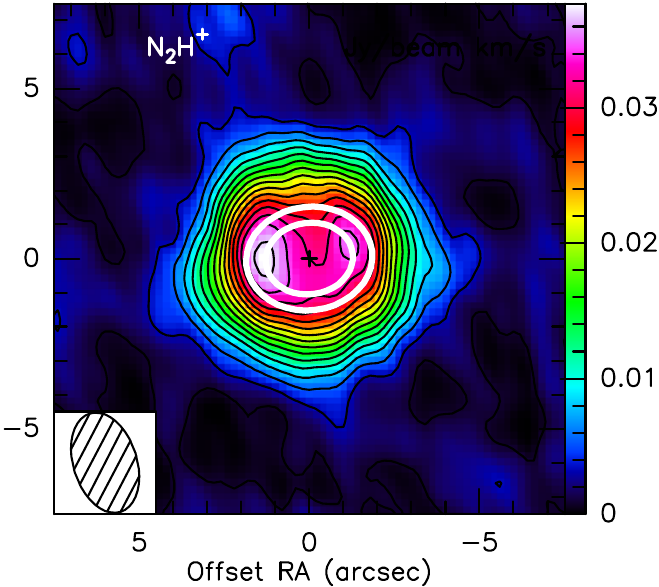}
 \includegraphics[width=0.22\linewidth]{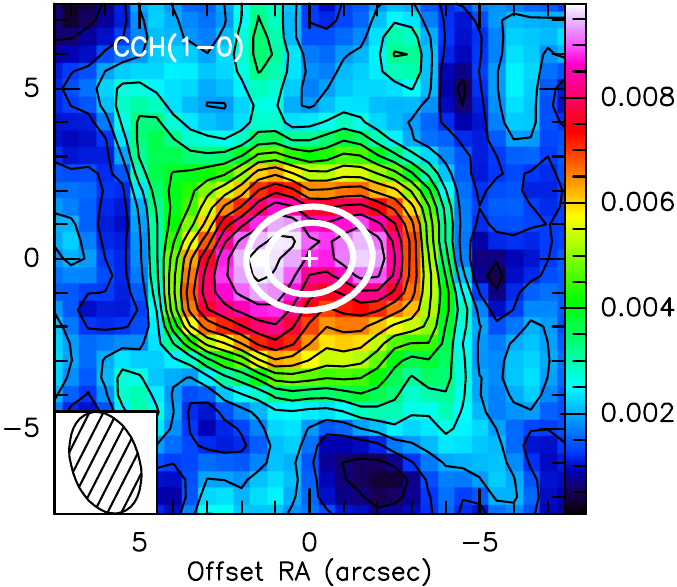}
\includegraphics[width=0.22\linewidth]{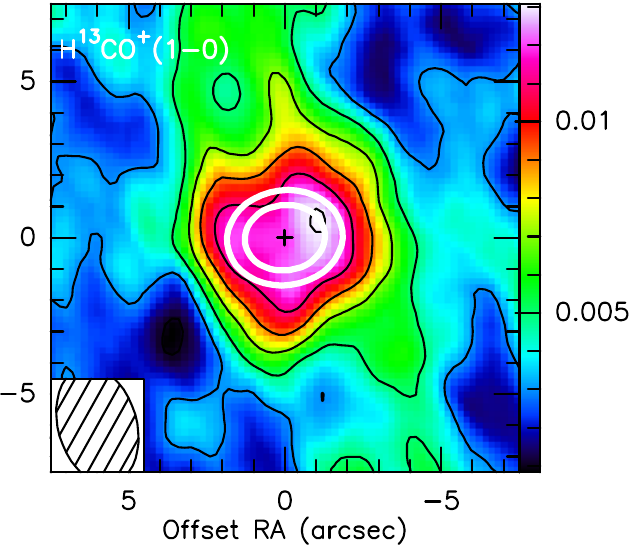}
\caption{\textit{From left to right, top to bottom:} Integrated intensity map of HCO$^+$(1--0), CS(2--1), CN(1--0),  HCN(1--0), HNC(1--0), N$_2$H$^+$, CCH, and H$^{13}$CO$^+$ (1--0). The white ellipses mark the approximate inner and outer edges of the dust ring at 180\,au and 260\,au, respectively. The beam size is indicated in the lower left corner and the colour scale is on the right is units of Jy\,beam$^{-1}$\,km\,s$^{-1}$. The intensity was integrated in the velocity range from 4.0 to 9.0\,km\,s$^{-1}$ with the threshold of 3$\sigma$. Contour levels are spaced by $5\sigma$ in the maps of HCO$^+$ and CS(2--1), and $3\sigma$ in the maps of CN, HCN, HNC, N$_2$H$^+$, CCH, and H$^{13}$CO$^+$ (1--0).} 
    \label{fig:highSNR}
    \end{figure*}

 \begin{figure}[h]
  \centering
\includegraphics[width=0.48\linewidth]{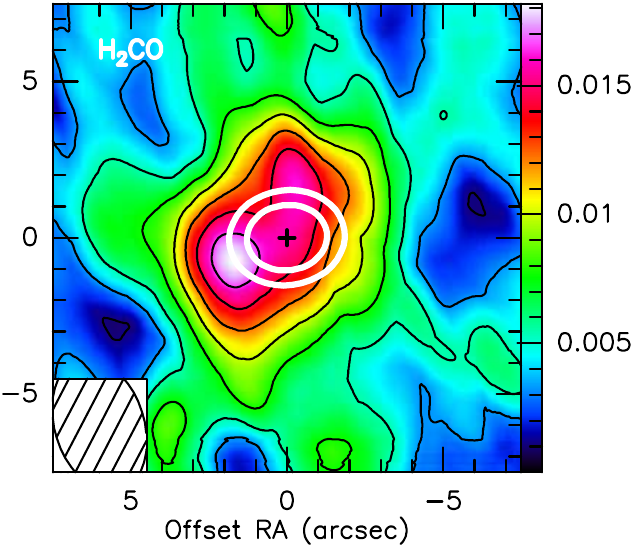}
\includegraphics[width=0.48\linewidth]{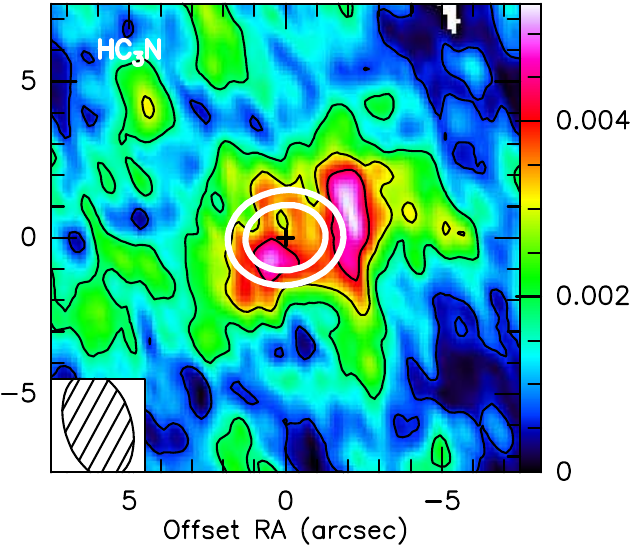}
 \caption{Same as Figure \ref{fig:highSNR}, but for H$_2$CO and HC$_3$N, respectively. The contour levels are spaced by $2\sigma$.} \label{fig:mediumSNR}
   \end{figure}
   
 \begin{figure*}[h]
 \centering
\includegraphics[width=0.25\linewidth]{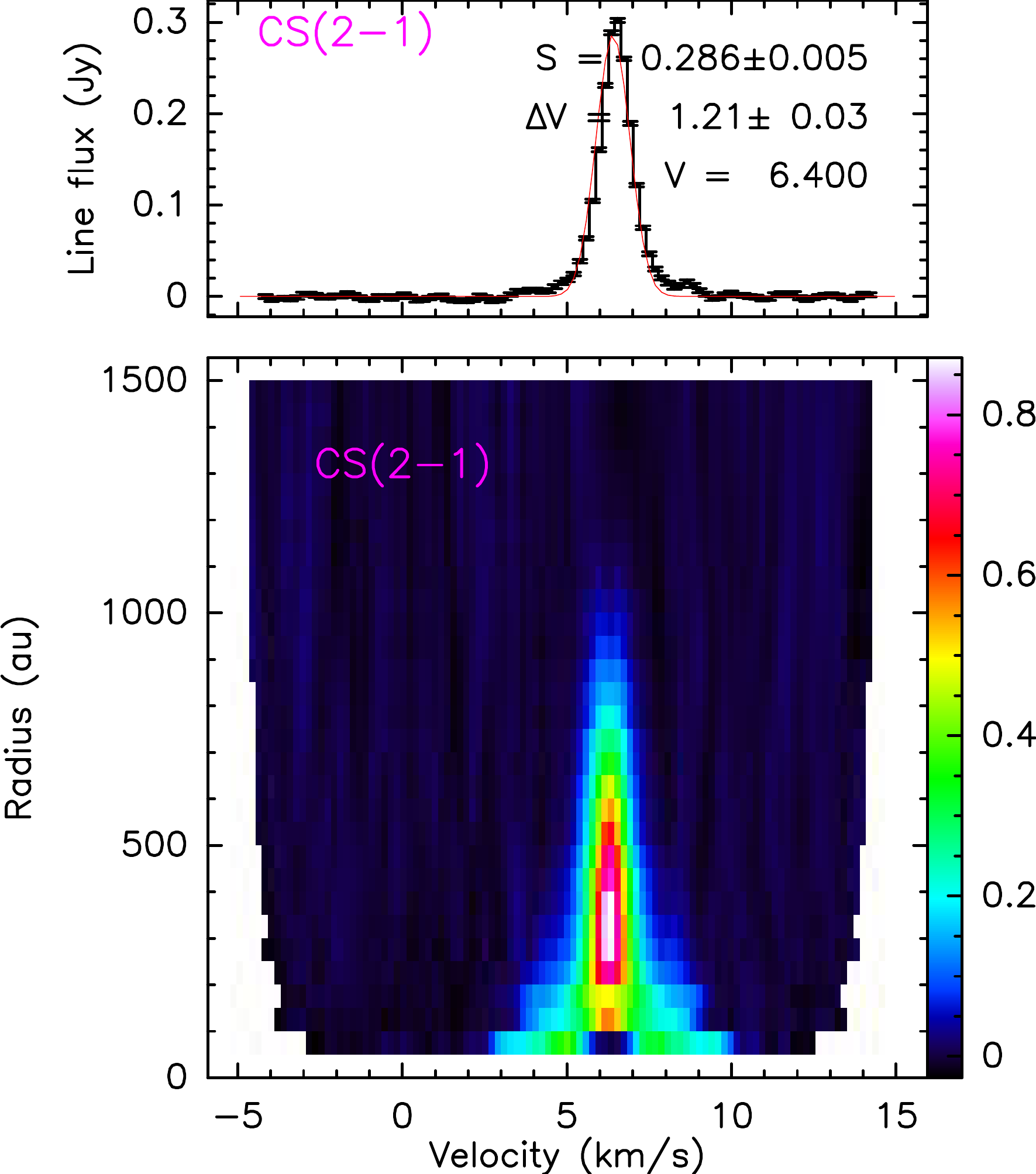}
\includegraphics[width=0.25\linewidth]{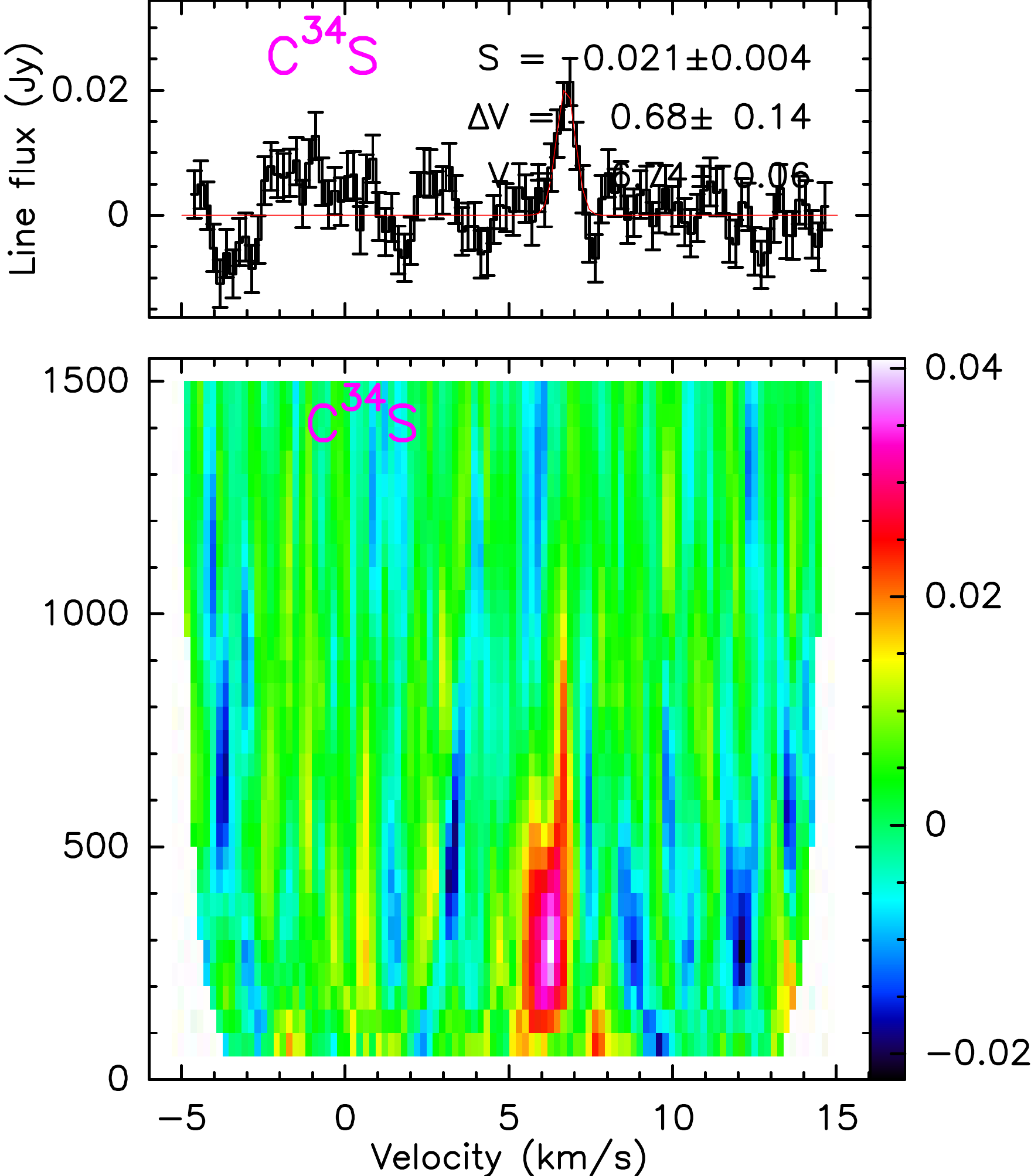}
 \includegraphics[width=0.25\linewidth]{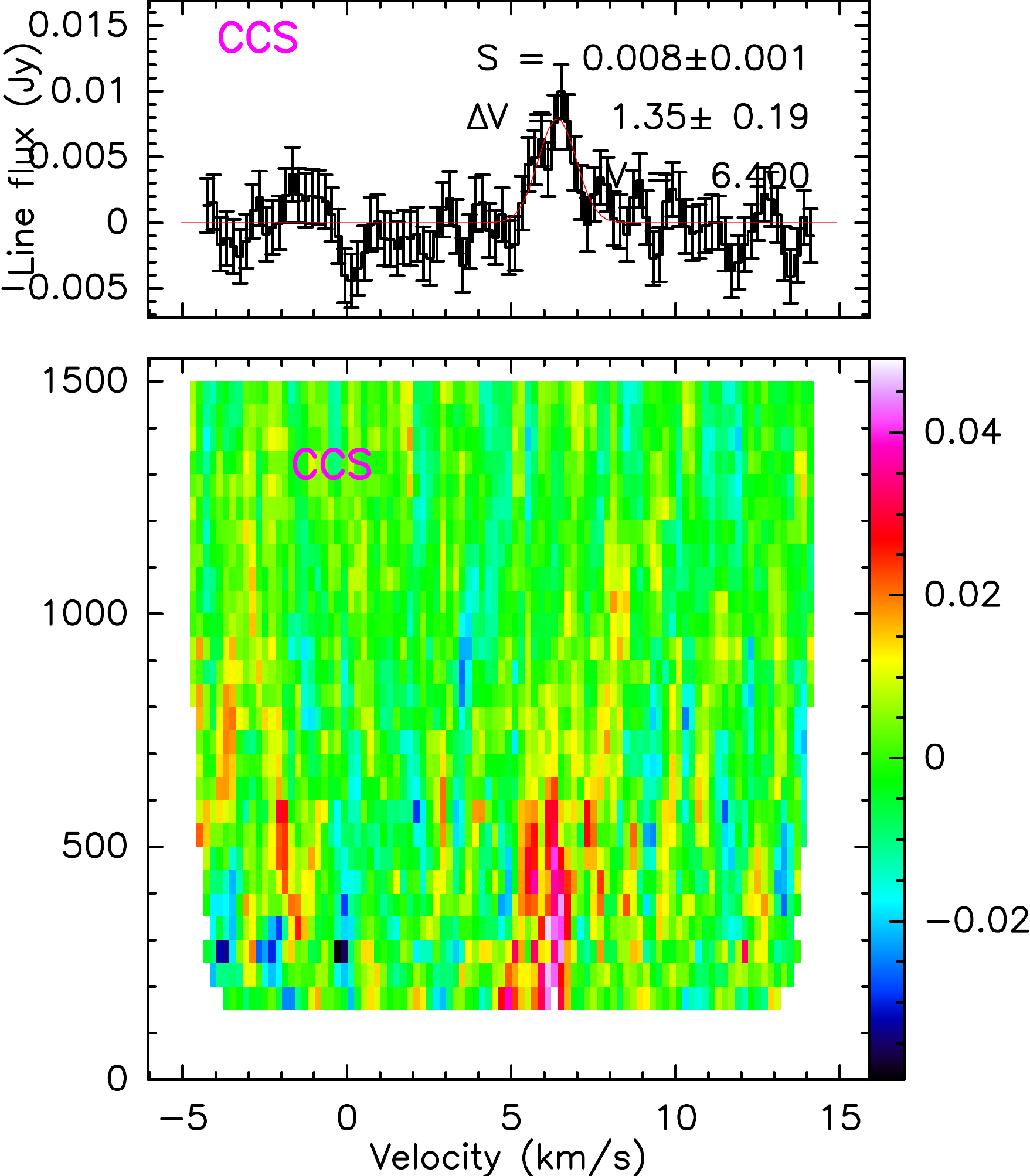}
  \includegraphics[width=0.26\linewidth]{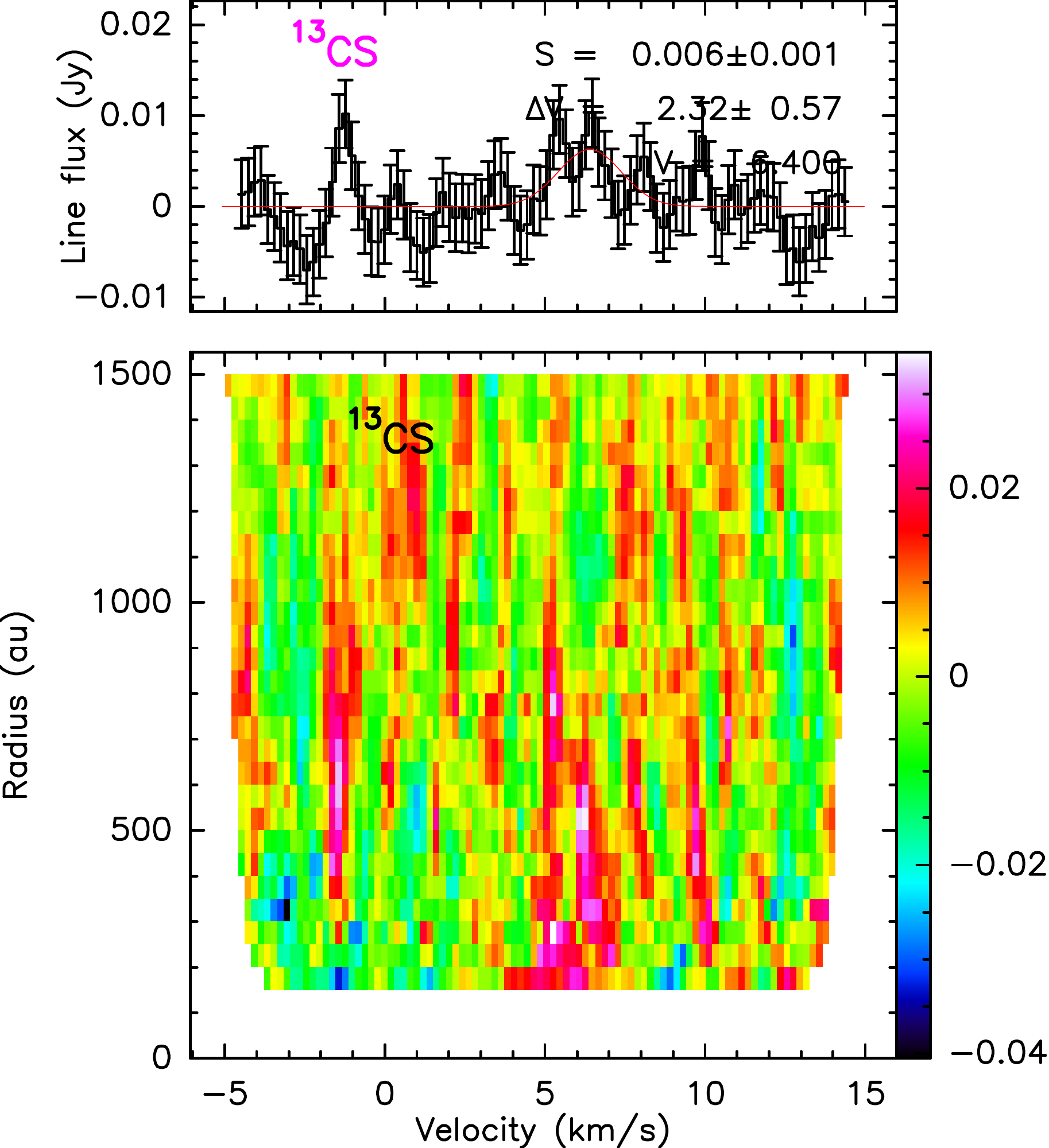}
  \includegraphics[width=0.25\linewidth]{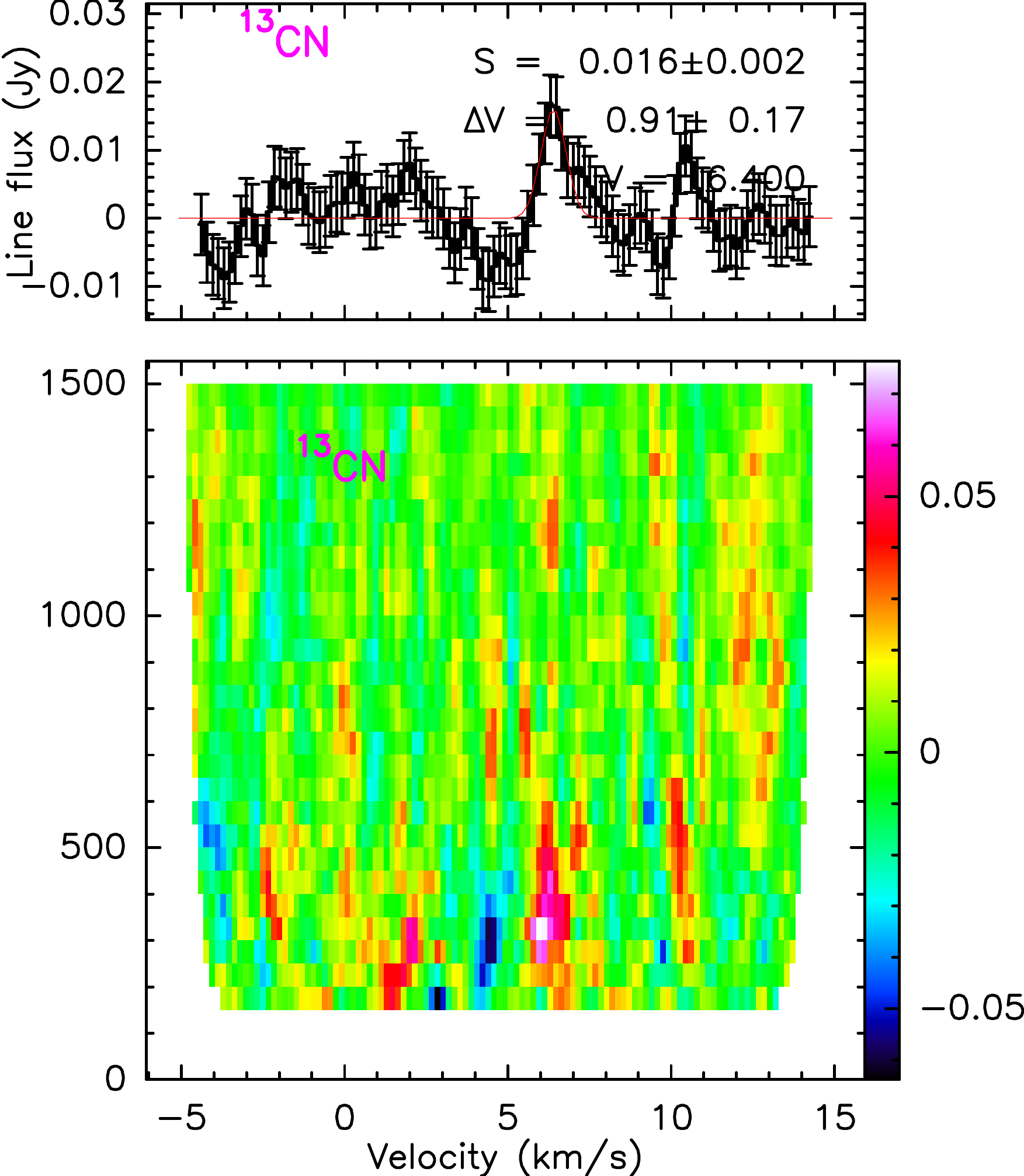}
 \includegraphics[width=0.25\linewidth]{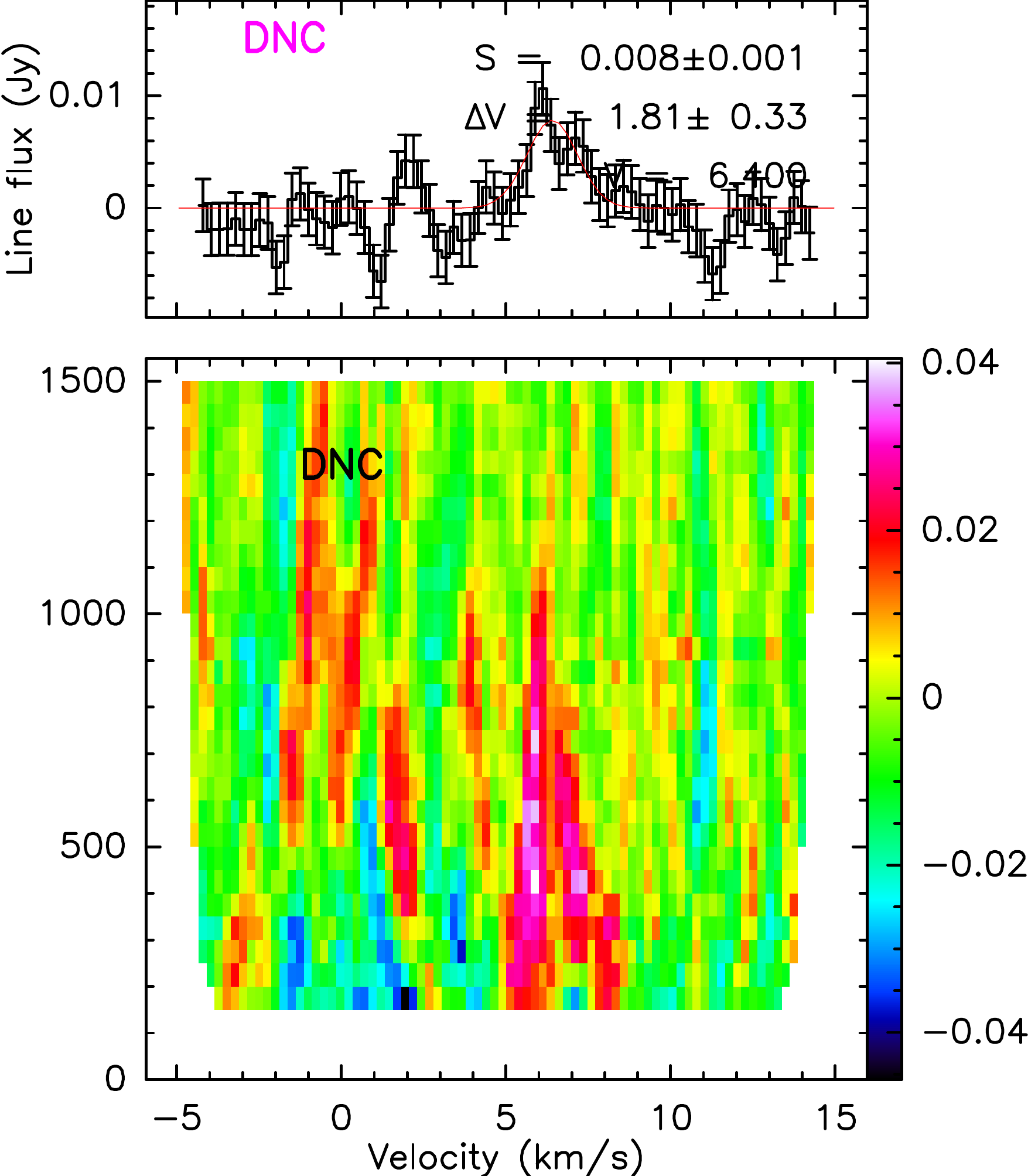}
    \caption{Integrated spectra (with Gaussian fits in red and fit results are indicated) and radial-velocity diagram of CS, C$^{34}$S(2--1), CCS,  $^{13}$CS(2--1), $^{13}$CN, and DNC obtained after making a Keplerian velocity correction.} \label{fig:lowSNR}
    \end{figure*}
    
 Among the 38 molecules which were observed in the survey, 17 are detected, including $^{13}$CO and C$^{18}$O, which we do not present in this paper. The high resolution maps of these two molecules can be found in \citet{Phuong+Dutrey+Diep_2020}.

The \textbf{CN, HCN, HNC, HCO$^+$, CS, N$_2$H$^+,$ and CCH} lines were detected at high S/N, as it can be seen in Fig.\ref{fig:highSNR}.
 All lines appear optically thin, as illustrated by the hyperfine ratios
of HCN, CN, CCH, and N$_2$H$^+$ (see Figure \ref{fig:rv}), and also by the peak brightness of HCO$^+$ (5.5 K).

Some molecules observed with a sufficiently high angular resolution (HCO$^+$, HCN, and N$_2$H$^+$)
display a clear east-west asymmetry. In a tilted Keplerian disk, the maximum opacity was obtained
along the major axis, resulting in two symmetric peaks. The observed asymmetry is most obvious
in HCO$^+$, suggesting that it is related to the 'hot spot' detected at PA 120$^\circ$ \citep{Dutrey+DiFolco+Guilloteau_2014, Tang+Dutrey+Guilloteau+etal_2016, Phuong+Dutrey+Diep_2020}. Furthermore, CCH may also be enhanced there, while the HNC map suggests a (marginal)
deficit at this position.

The apparent peak towards GG Tau for CN and CS emission is due to the lower angular resolution
of these data.\ Their RV diagrams (Figs. \ref{fig:lowSNR} and \ref{fig:rv}) clearly show that no emission is coming from inside about 200\,au.

As is shown in Figs.\ref{fig:highSNR}-\ref{fig:lowSNR}, \textbf{H$^{13}$CO$^+$, para-H$_2$CO, HC$_3$N, and C$^{34}$S} molecules have been detected with moderate S/N. The RV diagrams show that H$^{13}$CO$^+$ and N$_2$H$^+$ are more prominent in the dense ring, while CCH and H$_2$CO come only from the outer disk, beyond 300\,au out to about 600\,au (see Fig~\ref{fig:rv}). The non-detection of the o-H$_2$CO $6_{(1,5)}-6_{(1,6)}$ transition
is consistent with the expected temperature and an ortho-para ratio of 3.

The \textbf{DNC, $^{13}$CS, $^{13}$CN, and CCS} lines have been detected at low S/N ($\sim$6--8\,$\sigma$).
Since the data are very noisy, in Fig.\ref{fig:lowSNR}, we have chosen to only present their integrated spectra (\textit{top}) and the RV diagram (\textit{bottom}) obtained after Keplerian velocity correction, which best illustrate the detectability.

\textbf{Regarding the undetected molecules list,} we may have marginal detections ($\sim$2--3\,$\sigma$) of OCS, and perhaps DCO$^+$ and DCN.\ However, could not detect $^{13}$C$^{17}$O, N$_2$D$^+$, H$^{13}$CN, HC$^{15}$N, HN$^{13}$C, H$^{15}$NC, HOC$^+$, HCNH$^+$, HCCCHO, SO, SO$_2$, H$_2$CS, SiO, CCD, HDO, D$_2$CO, c-C$_3$H$_2$, or CH$_3$CN molecules.
\section{Radiative transfer modelling with DiskFit} 
\label{sec:data-ana}
The data were compared in the $uv$-plane with visibilities predicted for a disk model using the radiative transfer code DiskFit \citep{Pietu+Dutrey+Guilloteau+etal_2007}. A description of DiskFit usage for GG Tau can be found in the appendix of \citet{Phuong+Dutrey+Diep_2020}. 

The geometric parameters (inclination, orientation) and the physical power laws (velocity and temperature), derived from previous papers, are given in Table \ref{tab:disk}. As in \citet{Phuong+Chapillon+Majumdar_2018}, we kept the velocity and temperature laws as well as the power index ($p=1.5$) of the molecular surface density fixed. Only $\Sigma_{250}$, the value of the molecular surface density at 250\,au, was left free and varied during the minimisation process. 

We assumed that all molecules, except sulphur-bearing species, arise from the same layer as CO, so that they share the same temperature
profile, \mbox{$T(r)=25\,(r/200\,\textnormal{au})^{-1.0}$\,K}. 
Following \citet{Dutrey+Guilloteau+Pietu_2017} on the Flying Saucer edge-on disk, CS arises from the lower part of the molecular layer, closest to the disk midplane. We thus decided to analyse sulphur-bearing species with a lower temperature, namely the temperature profile derived from dust emission
at millimetre wavelengths \citep[$T(r)=15\,(r/200\,\mathrm{au})^{-1.0}$\,K,][]{Dutrey+DiFolco+Guilloteau_2014}. 
Because optically thin 1--0 line intensities scale roughly as $\Sigma/T$ above 10\,K,  the temperature
spread due to the specific molecular layer can only affect the derived surface density by a factor of 2 at most.  The factor is even smaller
for heavier molecules such as CCS, HC$_3$N, and OCS, since the observed lines are close to the brightest ones expected at these temperatures.
The molecular surface densities are given in Table \ref{table:cd}, with $3 \sigma$ upper limits for non-detections. Assuming an exponent $p=1$ only changes the surface densities by less than 30 \% (see Table \ref{tab:cd_1p0}).

\begin{table}[!th] 
\tiny 
\caption{GG Tau parameters}             
\label{tab:disk}      
\centering    
\begin{tabular}{|c  c| l |}        
\hline              
\multicolumn{2}{|c|}{Geometry} & \multicolumn{1}{c|}{Law}\\ 
\hline 
Inclination & 35$^\circ$ & $V(r)=3.4\,(\frac{r}{100\,\textnormal{au}})^{-0.5}$ \, (km\,s$^{-1}$) \\ 
Orientation & 7$^\circ$ & $T(r)=T_0\,(\frac{r}{200\,\textnormal{au}})^{-1.0}$\,\,\,\,\,\,\,\,\,\,\,\, (K)\\ 
Systemic velocity & 6.4 km s$^{-1}$ & $\Sigma(r)=\Sigma_{250}\,(\frac{r}{250\,\textnormal{au}})^{-1.5}$\, (cm$^{-2}$)  \\ 
\hline 
\end{tabular}
\vspace{-2.5ex}
\tablefoot{\tiny{The inner and outer radii are 180 and 600\,au for detected molecules, respectively, and 800\,au for undetected molecules.}}
\end{table}

\begin{table}
\tiny
\caption{Molecular surface density at 250\,au derived with DiskFit. The temperature uncertainty only affects the derived densities by factors smaller than 2.}        
\label{table:cd}      
\centering                          
\setlength\tabcolsep{2pt}
\begin{tabular}{|ccc||ccc|}          
\hline
Molecules &   & Surface density & Molecules & & Surface density \\  
                 &   & (cm$^{-2}$) &  & & (cm$^{-2}$) \\               
\hline 
 $^{13}$CO & D & $1.10\times10^{16}\,^{\ast}$& CS &  D & ($2.6\pm0.03) 10^{13}$  \\ 
 C$^{18}$O &  D & $2.42\times10^{15}\,^{\ast}$& $^{13}$CS &  D & $(2.6\pm0.7) 10^{11}$ \\ 
  $^{13}$C$^{17}$O &  U & $<1.2\,10^{13}$& CCS & D & $(1.5\pm0.2) 10^{12}$   \\ 
 CN  &  D & $(7.7\pm0.1)\,10^{13}$ &OCS& M & $(4.5 \pm 2.3)\,10^{10}$   \\ 
  $^{13}$CN & D & $(2.8\pm1.1)\,10^{12}$ &$p/o$-H$_2$CS & U & $<8.2\,10^{13}/5.3\,10^{12}$  \\
 CCH &  D & $(6.8\pm0.1)\,10^{13}$ & SO &   U  & $<4.5\,10^{12}$ \\
 N$_2$H$^+$ &  D & $(1.2\pm0.1)\,10^{12}$ &  SO$_2$ & U & $<5.0\,10^{12}$   \\
 N$_2$D$^+$ & U & $<1.3\,10^{11}$ & SiO & U & $<4.5\,10^{11}$\\
 HCN  &  D & $(6.70\pm0.04)\,10^{12}$ & DCN & M & $(1.9\pm1.2) 10^{11}$ \\ 
  H$^{13}$CN &  U  & $<1.6\,10^{11}$ & CCD&  U & $<1.0\,10^{14}$ \\
HC$^{15}$N &  U & $<2.9\,10^{11}$ & HDO& U & $<1.9\,10^{10}$ \\
  HNC  &  D  & $(3.4\pm0.03) 10^{12}$ & DNC &  D  & $(2.7\pm0.7)\,10^{11}$ \\ 
HN$^{13}$C & U &  $<1.7\,10^{11}$ & D$_2$CO & U & $<1.6\,10^{10}$ \\
  HCO$^+$  & D & $(1.50\pm0.01) 10^{13}$ & DCO$^+$  &M & $(2.2\pm0.7)\,10^{11}$   \\ 
  H$^{13}$CO$^+$ & D& $(4.0\pm0.2) 10^{11}$ & $p$-H$_2$CO & D & $(3.6\pm0.2)\,10^{12}$ \\
  HOC$^+$ & U & $<3.2\,10^{10}$ & c-C$_3$H$_2$&  U & $<1.0\,10^{12}$  \\      
HCNH$^+$ &  U & $<2.2\,10^{13}$  & HC$_3$N&   D & $(5.4\pm1.1) 10^{11}$ \\
HCCCHO  & U & $<1.4\,10^{17}$ &  CH$_3$CN & U & $<2.5\,10^{11}$\\

C$^{34}$S & D & $(1.0 \pm 0.1)\,10^{12}$ &  &   &  \\
 \hline                                  
\end{tabular}
\vspace{-2.5ex}
\tablefoot{\tiny{D=detected, U=undetected, and M=marginal detected\\
\tiny $T_0$ = 15\,K for S-bearing species, and $T_0$ = 25\,K for all other molecules \\
$^{\ast}$ The values are taken from \cite{Phuong+Dutrey+Diep_2020}, 
 }}

\end{table}
\section{Discussion}
\label{sec:dis}
\subsection{Sulphur in protoplanetary disk: First detection of CCS} 
Beyond CS, only a few S-bearing species observed in molecular clouds are detected in disks: H$_2$S \citep{Phuong+Chapillon+Majumdar_2018},
H$_2$CS \citep{LeGal+Oberg_2019}, SO \citep{Riviere-Marichalar+etal_2020}, and SO$_2$ \citep[although in a very atypical, warm disk,][]{Booth+Marel+Leemker_2021}.
In the GG Tau ring, we detected CCS, $^{13}$CS, and C$^{34}$S but failed to detect H$_2$CS, SO, and SO$_2$.
Though OCS may be marginally detected, however (see Appendix).

\begin{table}[h!]
\caption{Molecular abundance with respect to $^{13}$CO: $10^5\times(X_{mol}/X_{^{13}CO})$}    
\label{table:dens}      
\centering                          
\tiny
\setlength\tabcolsep{1.1pt}
\begin{tabular}{|lccc||lccc|}          
\hline
Mol. &  TMC-1 &  LkCa 15 & GG Tau & Mol. &  TMC-1 &  LkCa 15 & GG Tau \\                
\hline 
 C$^{18}$O & $1.1\,10^4\,^{(1)}$ & $2.8\,10^4\,^{(7)}$  & $2.2\, 10^{4}$ & C$^{34}$S &  ...  & ... & $ 10 \pm 1 $\\ 
\hline 
 CN  &  $2250\,^{(1)}$ & $3100\,^{(8)}$ & $660 \pm 30$ & CS & $1500^{(3)}$  & $520\,^{(8)}$ & $ 230\pm 10$\\ 
  $^{13}$CN & ... &  ... & $ 25 \pm 10$ &$^{13}$CS & $11\,^{(4)}$  & $2.8\,^{(10)}$ & $ 2.2\pm 0.6 $\\
  \cline{1-4}  
 CCH & $5960\,^{(2)}$ & $1200\,^{(8)}$& $600 \pm 30$ & CCS & $240\,^{(3)}$& ... & $ 13 \pm 2 $ \\
 N$_2$H$^+$ & $7680^{(1)}$  & $19.1\,^{(9)}$& $10.5 \pm 0.5$ &OCS & $1500\,^{(1)}$  & ... & $ 0.4 \pm 0.2 $ \\
  \hline  
 HCN  & $1500^{(2)}$ & $300\,^{(8)}$ & $57 \pm 3$ & DCN & $22\,^{(6)}$ & $7.5\,^{(9)}$ & $ 1.6 \pm 1.0$ \\ 
   HNC  & $1500^{(2)}$ & ...  & $29 \pm 2$  & DNC & $124\,^{(6)}$    & $3.5\,^{(9)}$ & $ 2.3 \pm 1.1$\\ 
   \hline  
  HCO$^+$  & $596\,^{(2)}$ & $350\,^{(8)}$ & $ 125 \pm 5$ & DCO$^+$  & $30\,^{(5)}$ & $4.5\,^{(11)}$ & $ 3.5 \pm 0.2\,^{(13)}$ \\ 
  H$^{13}$CO$^+$ &$8.3\,^{(1)}$ & $5.0\,^{(12)}$ & $3.4 \pm 0.2 $ & & & &\\
  \hline
  H$_2$CO & $1500\,^{(2)}$  & $13.6\,^{(9)}$  & $33 \pm 2$ (*) & HC$_3$N&  $473\,^{(2)}$   & $7.3\,^{(12)}$ & $ 4.6 \pm 0.9 $ \\
  \hline                                  
\end{tabular}
\vspace{-2.5ex}
\tablefoot{\tiny{ 
$^{(1)}$ \citet{Dutrey+Guilloteau+Guelin_1997}, 
$^{(2)}$ \citet{Omont_2007},
$^{(3)}$ \citet{Cernicharo+Cabezas_2021}
$^{(4)}$ \citet{Liszt+Ziurys_2012}, 
$^{(5)}$ \citet{Butner+Lada+Loren_1995},
$^{(6)}$ \citet{Turner_2001}, 
$^{(7)}$ \citet{Qi+Jacqueline_2003},
$^{(8)}$ \citet{Guilloteau+Reboussin+Dutrey+etal_2016},
$^{(9)}$ \citet{Loomis+Oberg+Andrews_2020},
$^{(10)}$ \citet{LeGal+Oberg_2019}, 
$^{(11)}$ \citet{Huang+Oberg+Qi+etal_2017}, 
$^{(12)}$ \citet{Chapillon+Dutrey+Guilloteau+etal_2012},
$^{(13)}$\citet{Phuong+Chapillon+Majumdar_2018}.
(*) for para-H$_2$CO only in GG Tau.
 Since estimating the uncertainties from all of these different studies was very difficult, we do not quote them for TMC1 and LkCa15.}} 
\end{table}
The CCS surface density measured in the GG Tau ring,  \mbox{$(1.5\pm0.2)\times10^{12}$\,cm$^{-2}$}, is of the order of 
the best previously reported upper limits in TTS and HAe disks.
Using the IRAM array, \citet{Chapillon+Dutrey+Guilloteau+etal_2012} reported upper limits (at 300\,au) of (0.9--1.4)$\times10^{12}$\,cm$^{-2}$ in LkCa\,15, GO\,Tau, DM Tau, and MWC 480 disks, while 
\citet{leGal_2021} through the ALMA Large Program MAPs, obtained
limits in the range \mbox{10$^{12}$--10$^{13}$\,cm$^{-2}$} for IM Lup, GM Aur, AS 209, HD 163296, and MWC 480 disks.
It is also consistent with our previous upper limit of \mbox{$< 1.7\times10^{12}$\,cm$^{-2}$} in GG Tau \citep{Phuong+Chapillon+Majumdar_2018}. 

Using Nautilus, a three phase gas-grain chemical model \citep{Ruaud+Wakelam+Hersant_2016}, \citet{Phuong+Chapillon+Majumdar_2018} modelled the chemistry in the GG Tau ring. They predicted a CCS surface density of $7.2\times10^{10}\,\mathrm{cm}^{-2}$ at 250 au, a factor 20 lower than the observed value. Our new upper limits of SO and SO$_2$ are comparable with the previous ones and are still in reasonable agreement with the chemical model presented in \citet{Phuong+Chapillon+Majumdar_2018}. This model is also compatible with our detection of HC$_3$N, which was not detected by \citet{Phuong+Chapillon+Majumdar_2018}.
However, for H$_2$S, \citet{Phuong+Chapillon+Majumdar_2018} predicted a surface density of $3.4\times10^{13} \mathrm{cm}^{-2}$ a factor 100 larger than what was observed. They conclude that H$_2$S is likely transformed in more complex species at grain surfaces, precluding any thermal desorption. The failure to predict CCS further supports the idea of an incomplete handling of the complex sulphur chemistry in disk chemical models, in addition to the difficulty of estimating the amount of sulphur depletion in refractory materials in grains prior to disk formation.

\subsection{ Comparison with TMC1 and LkCa15} 

Table \ref{table:dens} is a comparison of molecular abundances relative to $^{13}$CO with the TMC1 dark cloud and the transition disk \citep[cavity of radius 50\,au][]{Pietu+Dutrey+Guilloteau+etal_2007} of LkCa 15, a T Tauri star with a disk mass \mbox{$\sim$0.028\,M$_{\odot}$} \citep{Guilloteau+Dutrey+Pietu_2011}. 
For GG Tau A, we used a $^{13}$CO column density of \mbox{$\Sigma_{250}=1.1\times10^{16}$\,cm$^{-2}$}  \citep{Phuong+Dutrey+Diep_2020}, while for LkCa 15 we used \mbox{$\simeq 3.6\times10^{15}$\,cm$^{-2}$} \citep{Pietu+Dutrey+Guilloteau+etal_2007}. 

For TMC-1, we used \mbox{$\sim 1.1\times10^{16}$\,cm$^{-2}$}, following \citet{Cernicharo+etal_2021}. Relative abundances were then calculated using different studies.
At first order, these abundances (and the observed D/H ratios, see Appendix A) suggest that the GG Tau and LkCa\,15 disks share similar chemical properties. We note however that the radical abundances (CN and CCH) are somewhat lower towards GG Tau, perhaps because the GG Tau dense ring is shadowing the outer disk, reducing the UV flux. 

\textbf{Sulphur-bearing species}
The CS/C$^{34}$S ratio, $25\pm3$, is compatible with solar isotopic ratios and optically thin CS (2--1) emission.
Despite limited S/N, the CCS emission  (Fig.\ref{fig:lowSNR}) seems to arise from the dense ring, as for H$_2$S \citep{Phuong+Chapillon+Majumdar_2018}. 
We find that CS/CCS $\sim$17 is similar to the CS/H$_2$S ratio \citep{Phuong+Chapillon+Majumdar_2018}, and somewhat higher than the value reported in TMC1 
\mbox{\citep[$\sim $ 6.3,][]{Cernicharo+etal_2021}}. Moreover, we observe $^{13}$CS/$^{12}$CS $\simeq 0.01$, while \citet{LeGal+Oberg_2019} found 0.015 in LkCa 15.

\textbf{CCS can form via three routes:} The first route being \mbox{S$^+$+C$_2$H$_2$ $\rightarrow$ HCCS$^+$+H} and \mbox{HCCS$^+$+e $\rightarrow$ CCS+H} and the second one being \mbox{CCH+S $\rightarrow$ CCS+H}, which are exothermic and free of reaction barriers,
play important roles in the sulphur chemistry in molecular clouds, while the third route \mbox{CH+CS $\rightarrow$ CCS+H} is the main production pathway. \citet{Sakai+Ikeda+Morita_2007} reported  $^{13}$CCS/C$^{13}$CS$\sim$ 4 in TMC-1 and L1521E, indicating that the two carbon atoms making up CCS are not equivalent and have to come from different molecular species or non-equivalent positions of a single molecule  \citep[e.g. HC$_3$N, CCH, $c$-C$_3$H$_2$, C$_3$S, and C$_4$H, ][]{Takano+Suzuki+Ohishi_1990,Sakai+Saruwatari+Sakai_2010,Sakai+Shiino+Hirota_2010,Yoshida+Sakai+Tokudome_2015}. After formation, CCS can be a precursor of CS via several reactions \citep{Vastel+Quenard+LeGal_2018,Semenov+Favre+Fedele_2018}.
In the outer region of a protoplanetary disk, the sulphur chemistry starts with the S$^+$ ion, the least efficient formation route of CCS. This probably makes the molecule in PPDs much less abundant than in molecular clouds. 

\vspace{0.25cm}

The angular resolution of these data is insufficient for proper comparisons with a chemical model and the determination of radial variations, although physical conditions in the dense ring and the outer disk are known 
to be different \citep{Phuong+Dutrey+Diep_2020}.  Such comparisons would require high angular resolution images, but this is  a challenging task given the low frequencies at which heavy molecules such as CCS or OCS have their peak emission at temperatures of 15--25\,K or below. Nevertheless, the detection of a new rare species, CCS, in the dense ring of GG Tau confirms it is a good laboratory to study the cold PPD chemistry, thanks to its large size and mass.
 \begin{acknowledgements}
This work is based on observations carried out with the IRAM NOEMA Interferometer. IRAM is supported by INSU/CNRS (France), MPG (Germany) and IGN (Spain).
A. Dutrey and S. Guilloteau thank the French CNRS programs PNP, PNPS and PCMI. N.T.Phuong and P.N.Diep acknowledge financial support from World Laboratory, Rencontres du Viet Nam, and Vietnam National Space Center.  N.T.Phuong thanks financial support from Korea Astronomy and Space Science Institute. This research is
funded by Vietnam National Foundation for Science and Technology Development (NAFOSTED) under grant number 103.99-2019.368. A. C. acknowledges financial support from the Agence Nationale de la Recherche (grant ANR-19-ERC7-0001-01). C.W.L. is supported by the Basic Science Research Program through the National Research Foundation of Korea (NRF) funded by the Ministry of Education, Science and Technology (NRF-2019R1A2C1010851).
\end{acknowledgements}

\bibliography{references}
\bibliographystyle{aa}

%
\begin{appendix}
\section{Deuterated species}
\begin{figure}[h!]
 \centering  
\includegraphics[width=0.95\linewidth]{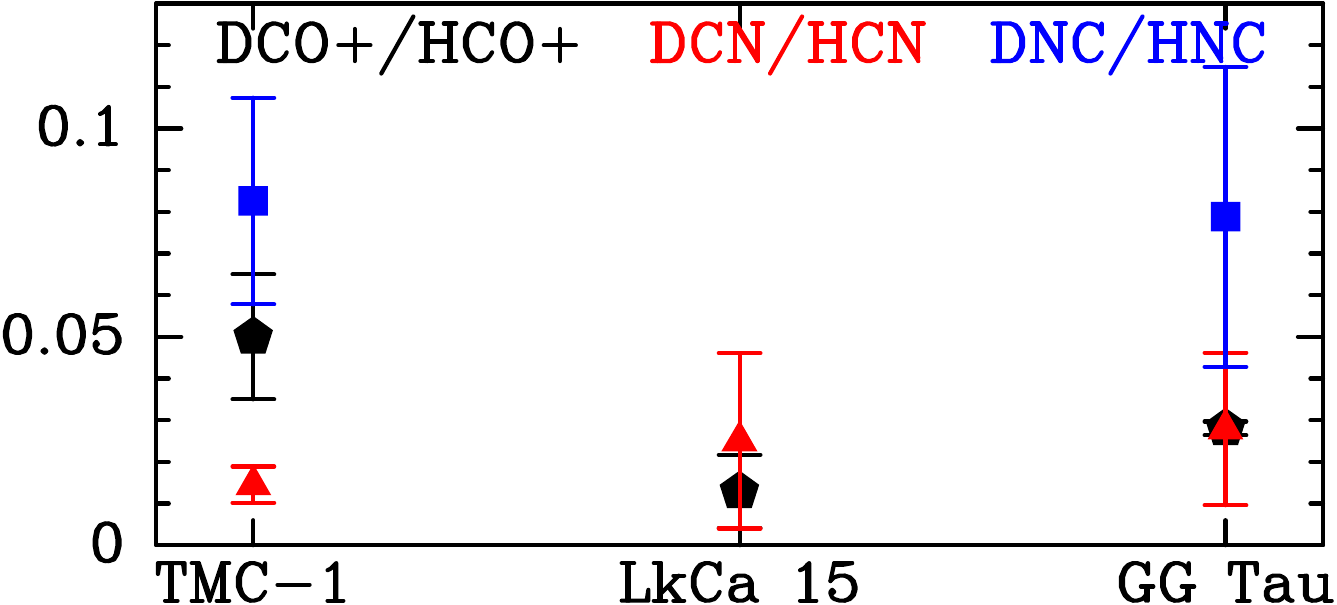}
\vspace{2ex}
  \caption{D/H ratios in TMC-1, LkCa 15, and GG Tau A.} \label{fig:d_over_h}
\end{figure}

All D/H ratios (Fig.\ref{fig:d_over_h}) are very similar (within the error bars) and around 0.01-0.05. As already quoted by several studies, such values are in agreement with dark cloud measurements and consistent with a low temperature chemistry \citep[e.g.][]{Phuong+Chapillon+Majumdar_2018, vanDishoeck+Thi+vanZadeldoff+etal_2003,Bergin+Cleeves+Gorti_2013,Dutrey+Semenov+Chapillon_2014,Huang+Oberg+Qi+etal_2017, Aikawa+Furuya+Hincelin_2018}. 

\section{Molecular surface density derived with power index of $p=1.0$ for surface density law}

\begin{table}[h!]
\caption{Molecular surface density at 250\,au 
assuming $p=1$.}     
\vspace{2ex}
\label{tab:cd_1p0}      
\centering                          
\tiny
\setlength\tabcolsep{2pt}
\begin{tabular}{|ccc||ccc|}          
\hline
Molecules &   & Surface density & Molecules & & Surface density \\  
                 &   & (cm$^{-2}$) &  & & (cm$^{-2}$) \\               
\hline 
 $^{13}$CO & D & $1.10\times10^{16}\,^{\ast}$& CS &  D & $(1.97\pm0.02)\,10^{13}$ \\ 
 C$^{18}$O &  D & $2.42\times10^{15}\,^{\ast}$& $^{13}$CS &  D & $(2.0\pm0.5)\,10^{11}$  \\ 
  $^{13}$C$^{17}$O &  U & $<2.1\,10^{13}$ & CCS & D & $(9.0\pm0.5)\,10^{11}$\\ 
 CN  &  D & $(7.0\pm0.2)\,10^{13}$ &OCS& M & $(1.3\pm0.4)\,10^{11}$    \\ 
  $^{13}$CN & D & $(2.1\pm0.8)\,10^{12}$  &$p/o$-H$_2$CS & U &  $<7.3\,10^{13}/3.7\,10^{12}$ \\
 CCH &  D & $(5.0\pm0.1)\,10^{13}$  & SO &   U  & $<2.8\,10^{12}$  \\
 N$_2$H$^+$ &  D & $(8.4\pm0.7)\,10^{11}$ &  SO$_2$ & U & $<1.9\,10^{12}$   \\
 N$_2$D$^+$ & U & $<2.0\,10^10$ & SiO & U & $<3.3\,10^{11}$ \\
 HCN  &  D & $(4.89\pm0.03)\,10^{12}$   & DCN & M & $(1.2\pm0.8)\,10^{11}$ \\ 
  H$^{13}$CN &  U  & $<1.1\,10^{11}$ & CCD&  U & $<7.35\,10^{13}$ \\
HC$^{15}$N &  U & $<2.0\,10^{11}$ & HDO& U & $<2.2\,10^{11}$ \\
  HNC  &  D  & $(2.47\pm0.02)\,10^{12}$  & DNC &  D  & $(2.0\pm0.5)\,10^{11}$  \\ 
HN$^{13}$C & U & $<1.3\,10^{11}$  & D$_2$CO & U & $<2.5\,10^{9}$  \\
  HCO$^+$  & D & $(1.10\pm0.03)\,10^{13}$ & DCO$^+$  &M & $(1.6\pm0.5)\,10^{11}$   \\ 
  H$^{13}$CO$^+$ & D& $(2.8\pm0.2)\,10^{11}$ & $p$-H$_2$CO & D & $(2.6\pm0.1)\,10^{12}$ \\
  HOC$^+$ & U & $<1.9\,10^{10}$ & c-C$_3$H$_2$&  U & $<9.7\,10^{11}$  \\      
HCNH$^+$ &  U & $<1.5\,10^{13}$  & HC$_3$N&   D & $(3.5\pm1.2)\,10^{11}$ \\
HCCCHO  & U & $<1.3\,10^{17}$ &  CH$_3$CN & U & $<2.2\,10^{11}$\\

C$^{34}$S & D & $(7.7\pm0.8)\,10^{11}$ &  &   &  \\
 \hline                                  
\end{tabular}
\tablefoot{\tiny{D=detected, U=undetected, and M=marginal detected\\
\tiny $T_0$ = 15\,K for S-bearing species, and $T_0$ = 25\,K for all other molecules \\
$^{\ast}$ The values are taken from \cite{Phuong+Dutrey+Diep_2020}. 
 }}
\end{table}
 \section{Radial profiles}
 \label{sec:prof}
\begin{figure}[h!]
 \centering  
 \includegraphics[width=0.75\linewidth]{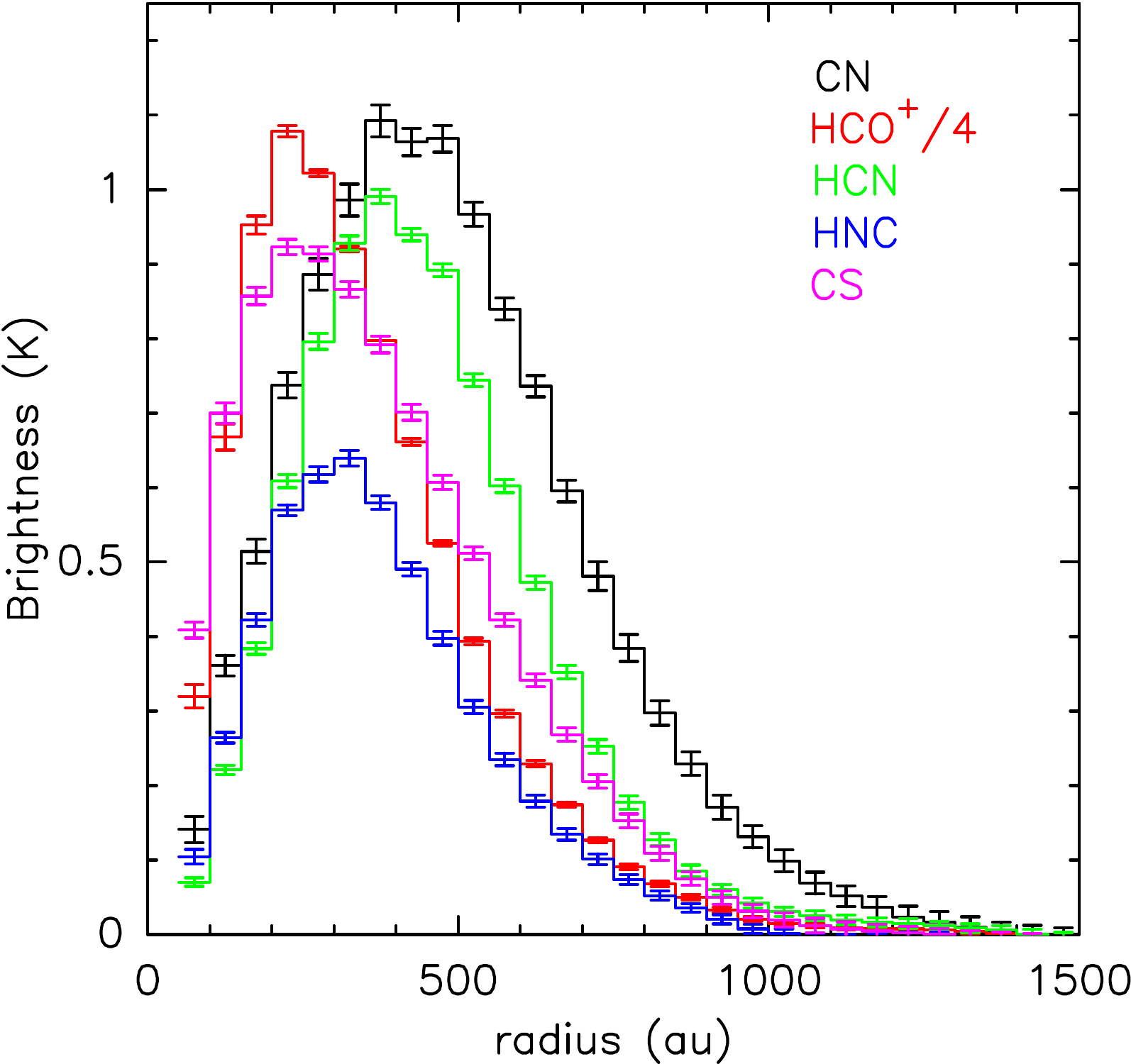}
  \includegraphics[width=0.75\linewidth]{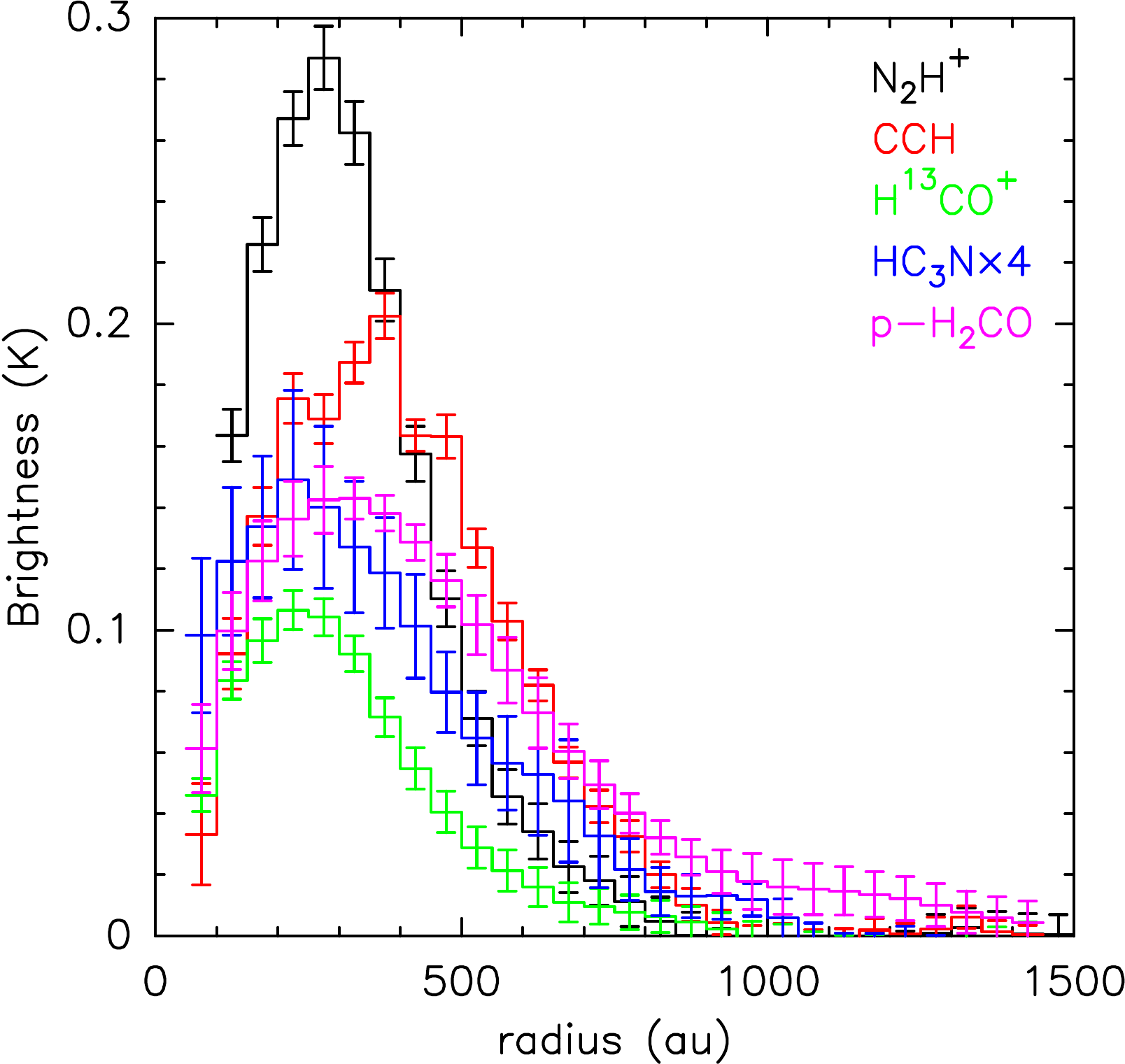}
  \caption{Radial profiles of brightness temperature at systemic velocity of 6.4\,km\,s$^{-1}$ of molecular lines detected at SNR$>5$.}.
\end{figure}
\section{Radial-velocity diagram of OCS}
\begin{figure}[h!]
\centering
 \includegraphics[width=0.85\linewidth]{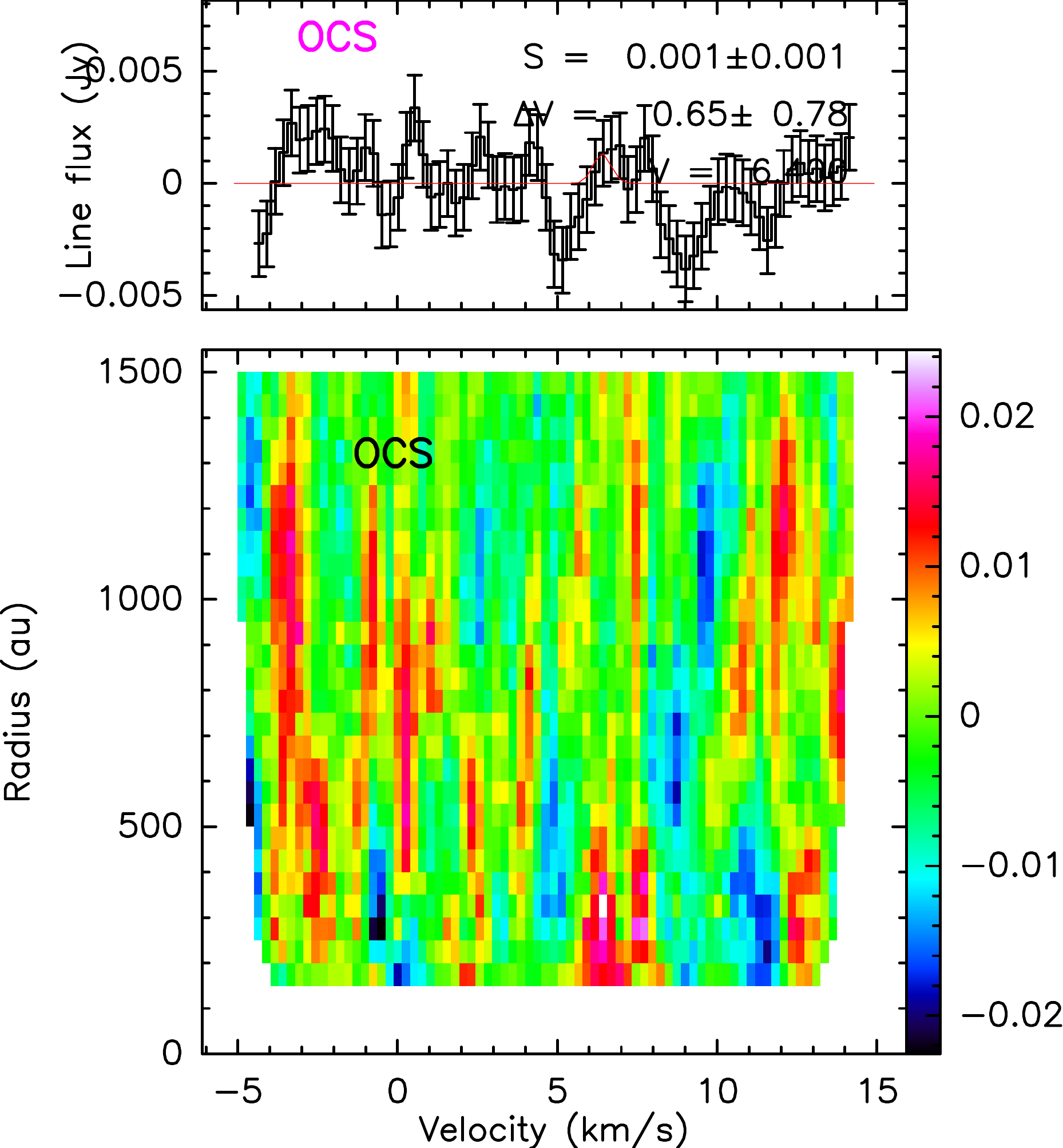}
 \vspace{2ex}
 \caption{Integrated spectra \textit{(top)} and radial-velocity diagram \textit{(bottom)} of OCS obtained after making a Keplerian velocity correction and line stacking.}
 \end{figure}
\section{Radial-velocity diagrams of other detected molecules}
 \begin{figure*}
\centering
\vspace{3ex}
\includegraphics[width=0.28\linewidth,trim=0cm 2cm 0cm 0cm]{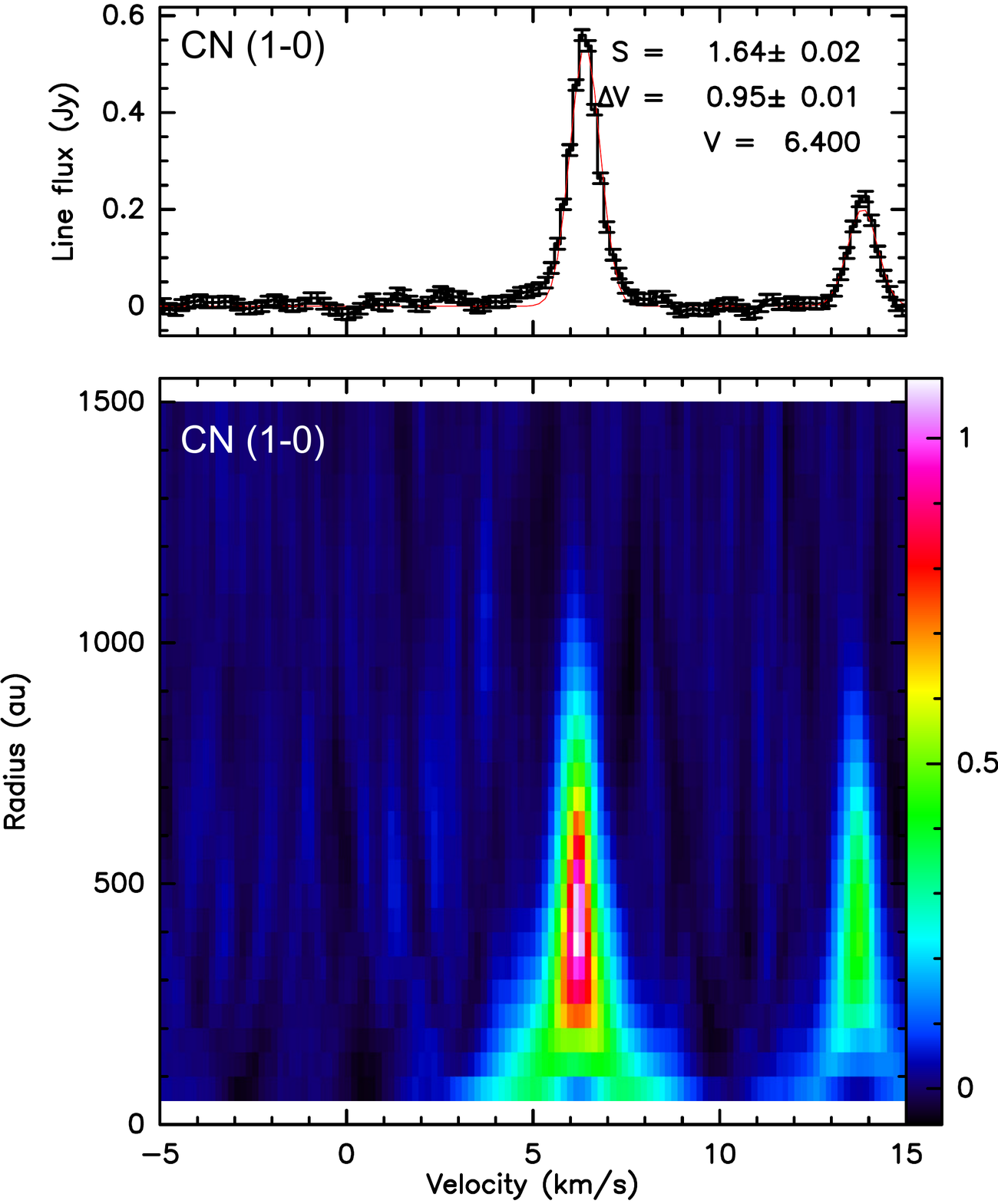}
\includegraphics[width=0.275\linewidth]{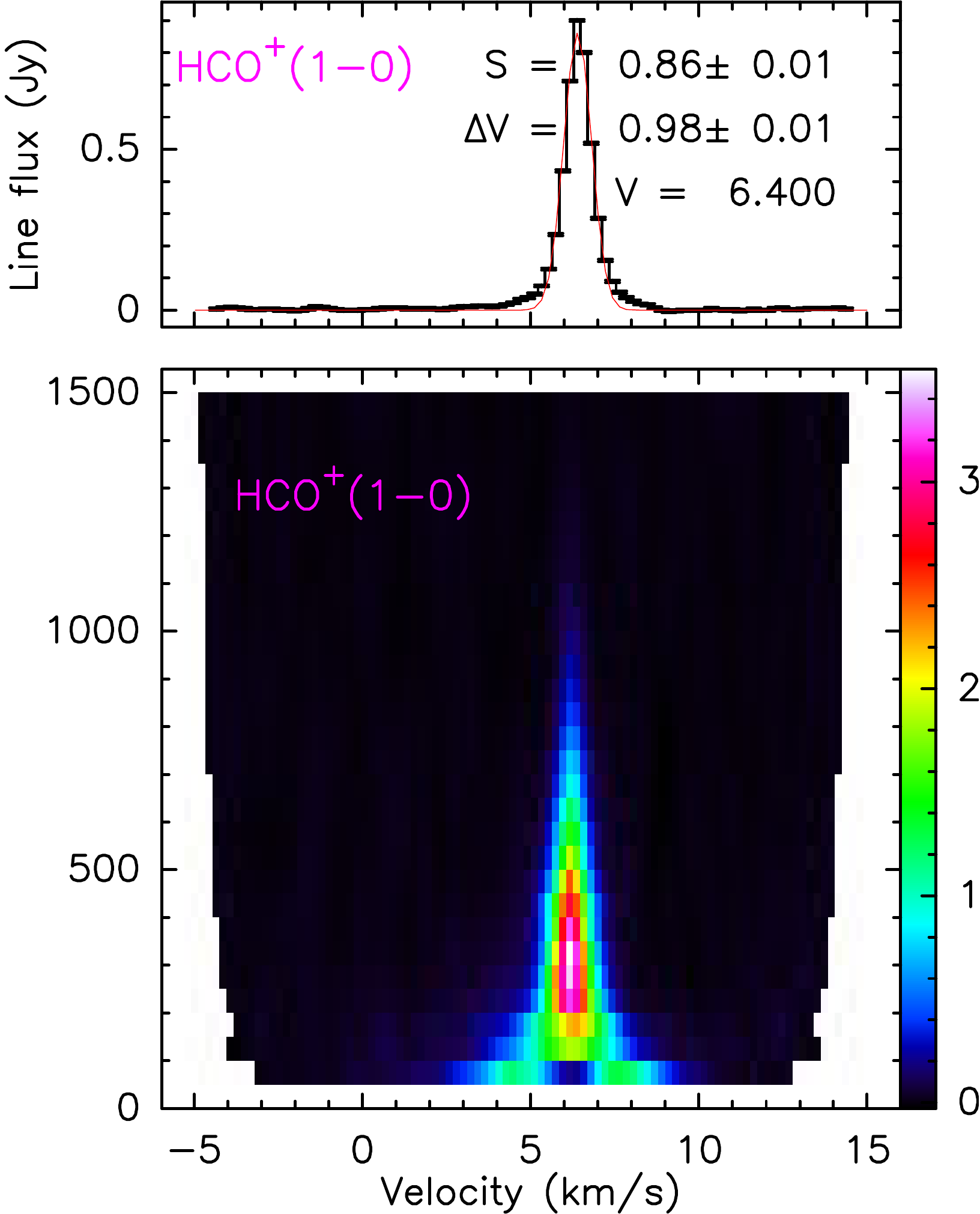}
 \includegraphics[width=0.27\linewidth]{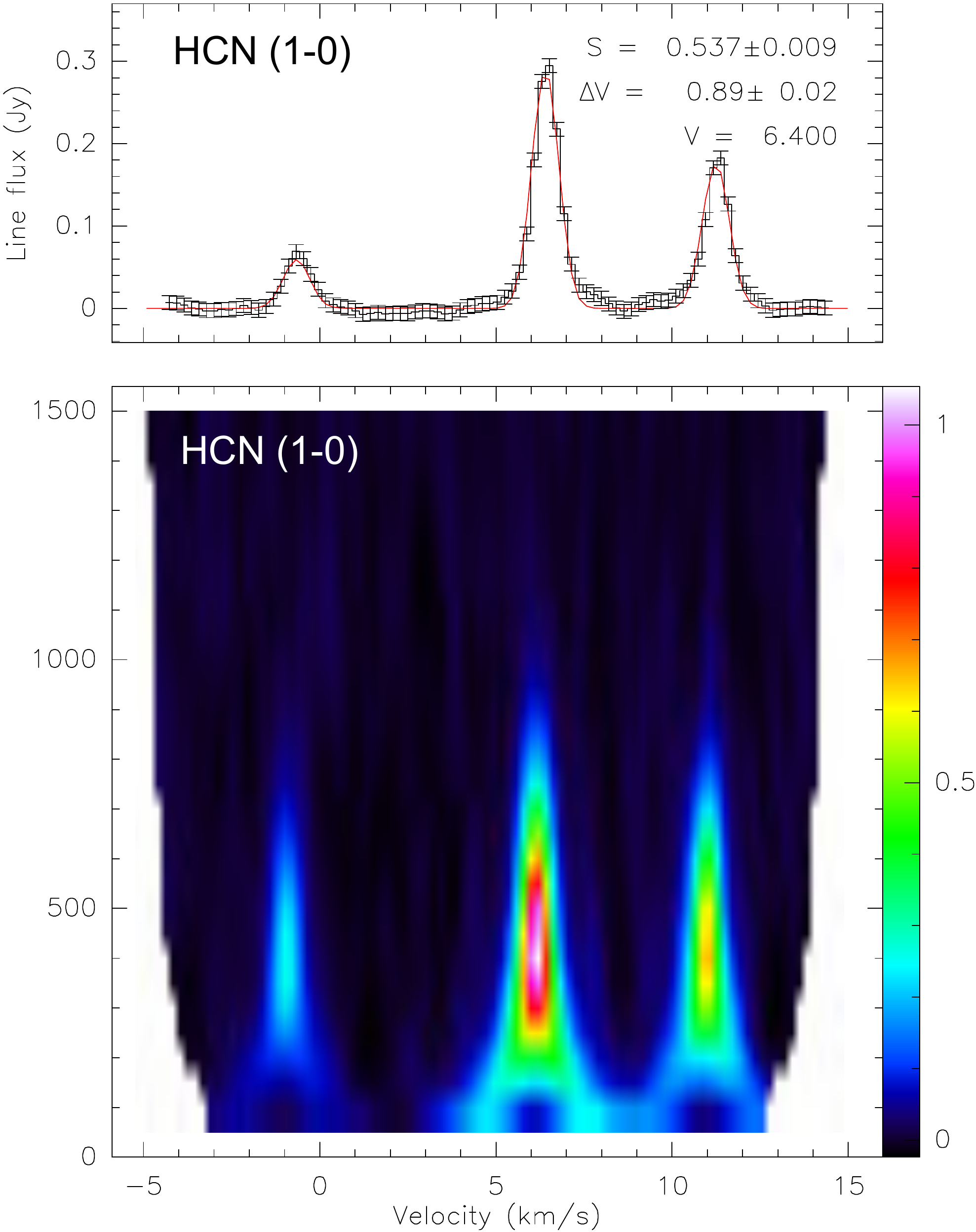}\\
 \vspace{3ex}
 \hspace{0.5cm} \includegraphics[width=0.29\linewidth]{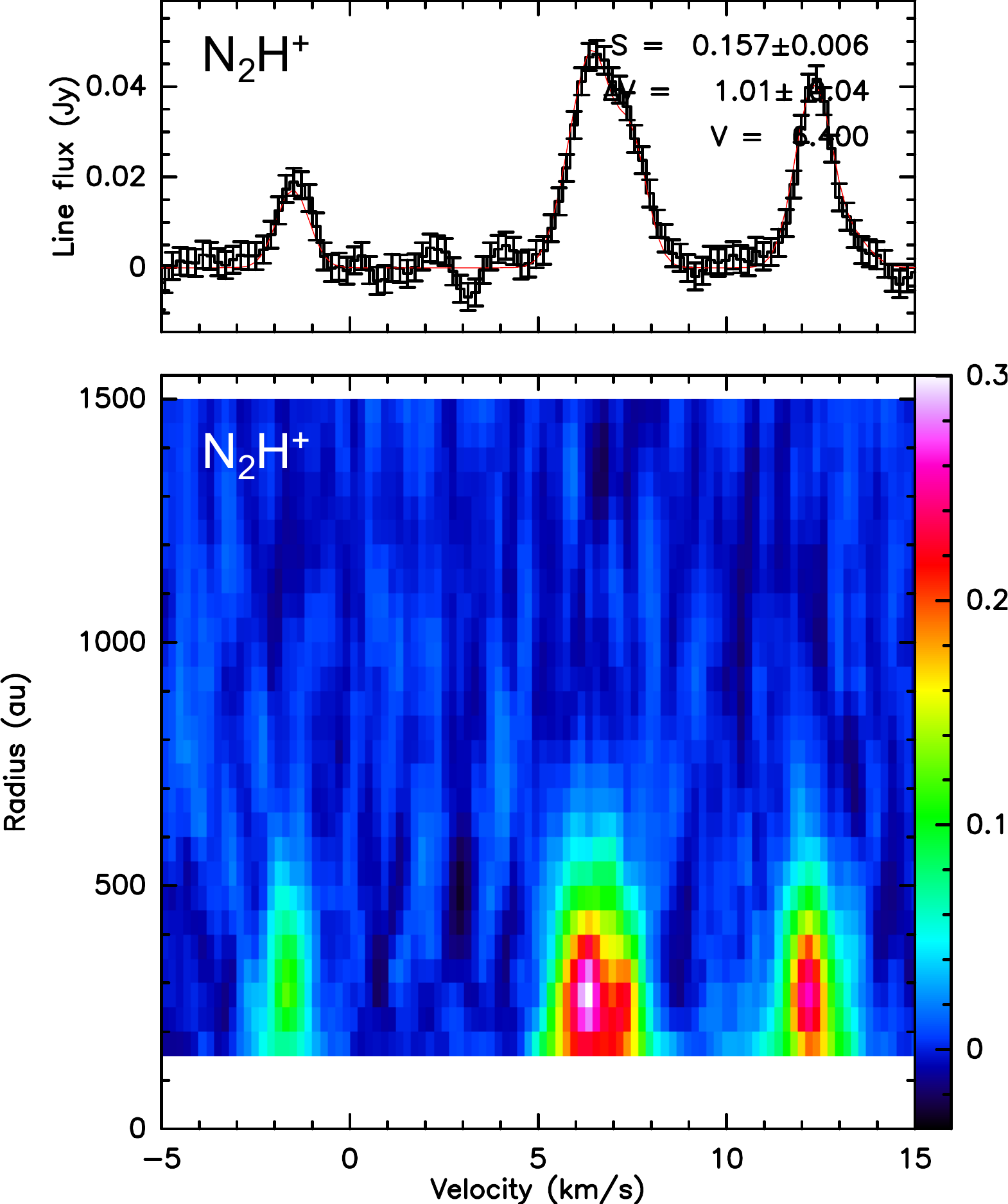}
 \includegraphics[width=0.285\linewidth]{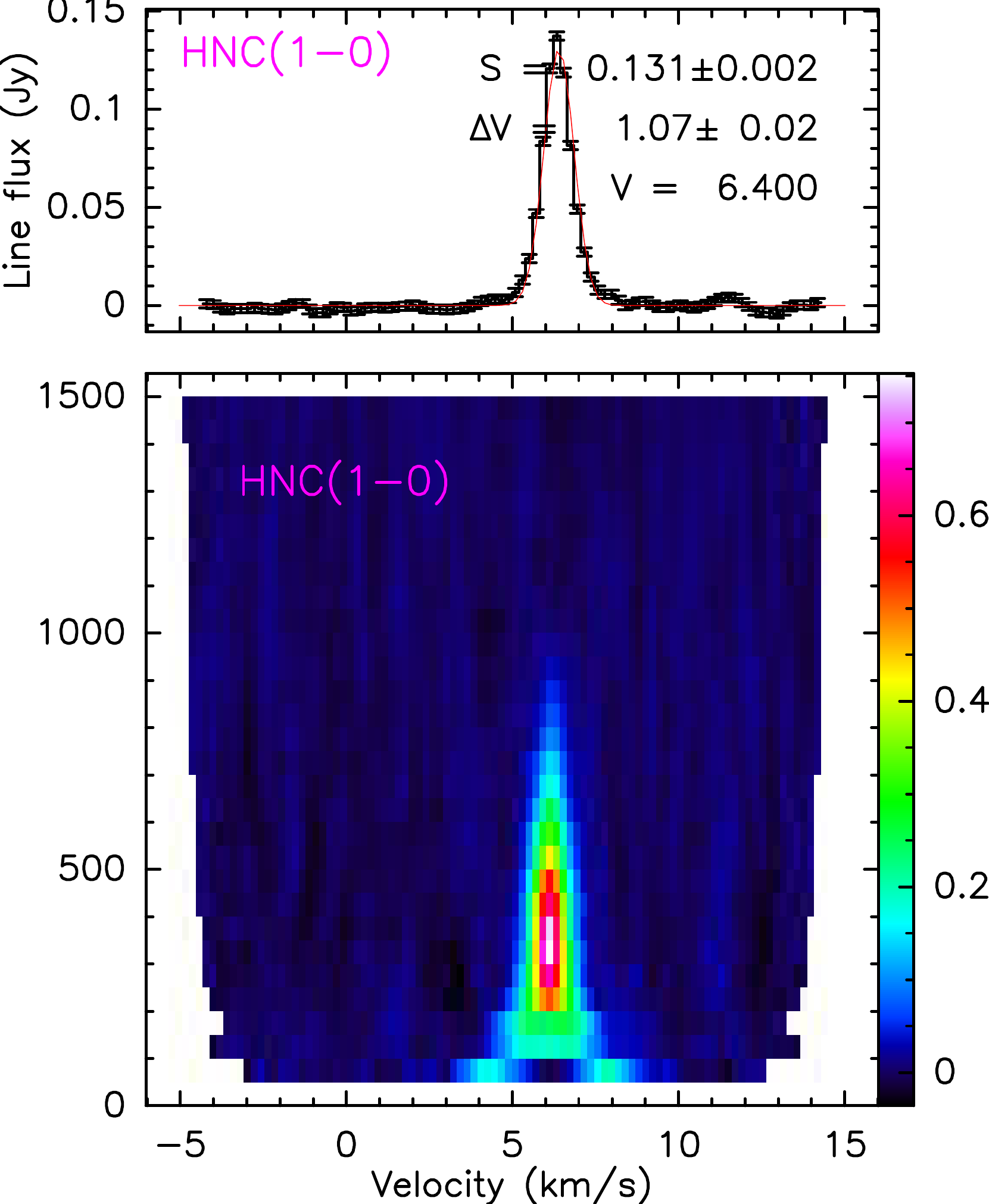}
 \includegraphics[width=0.29\linewidth]{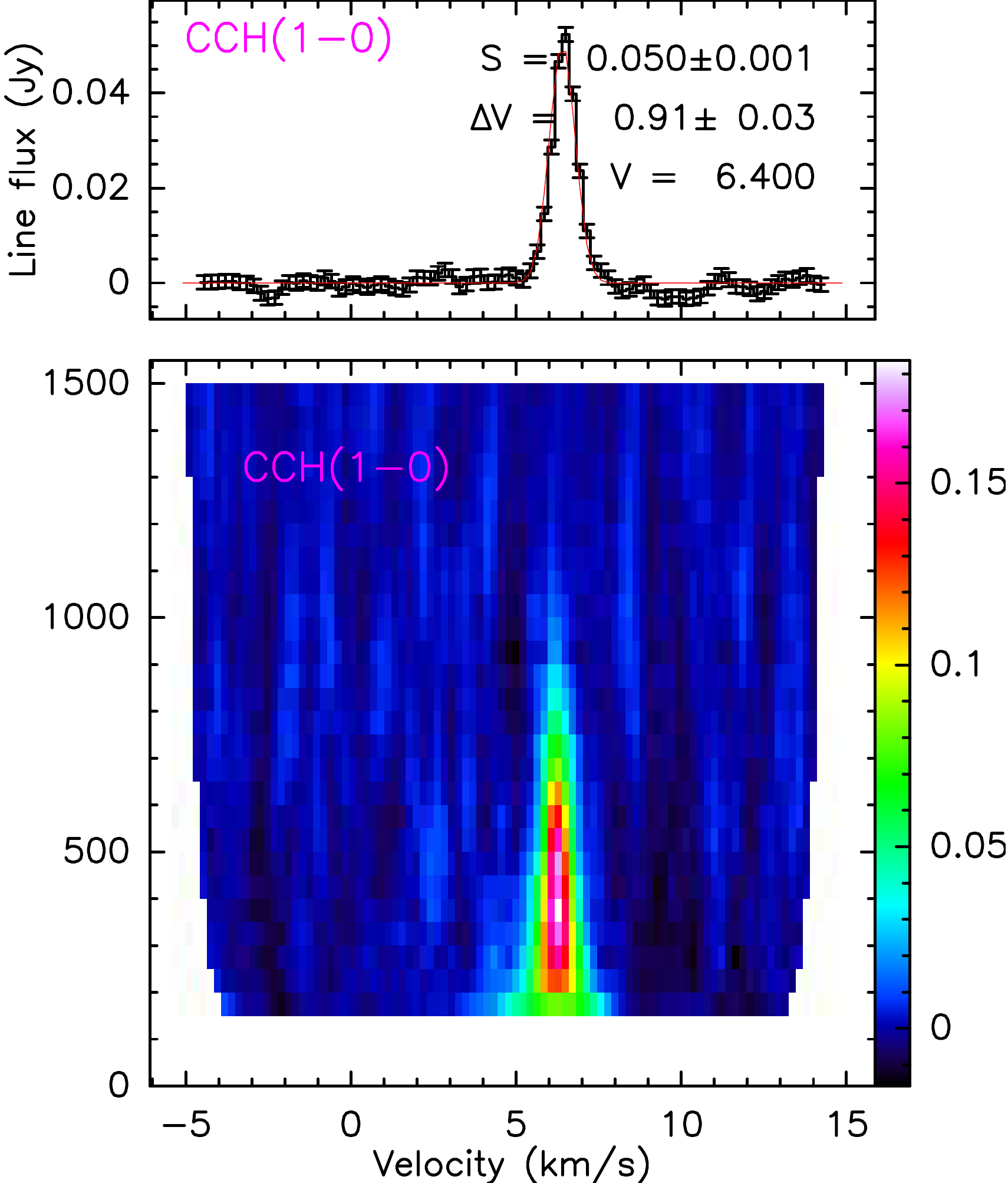}\\
  \vspace{2ex}
 \hspace{1.4cm}\includegraphics[width=0.3\linewidth]{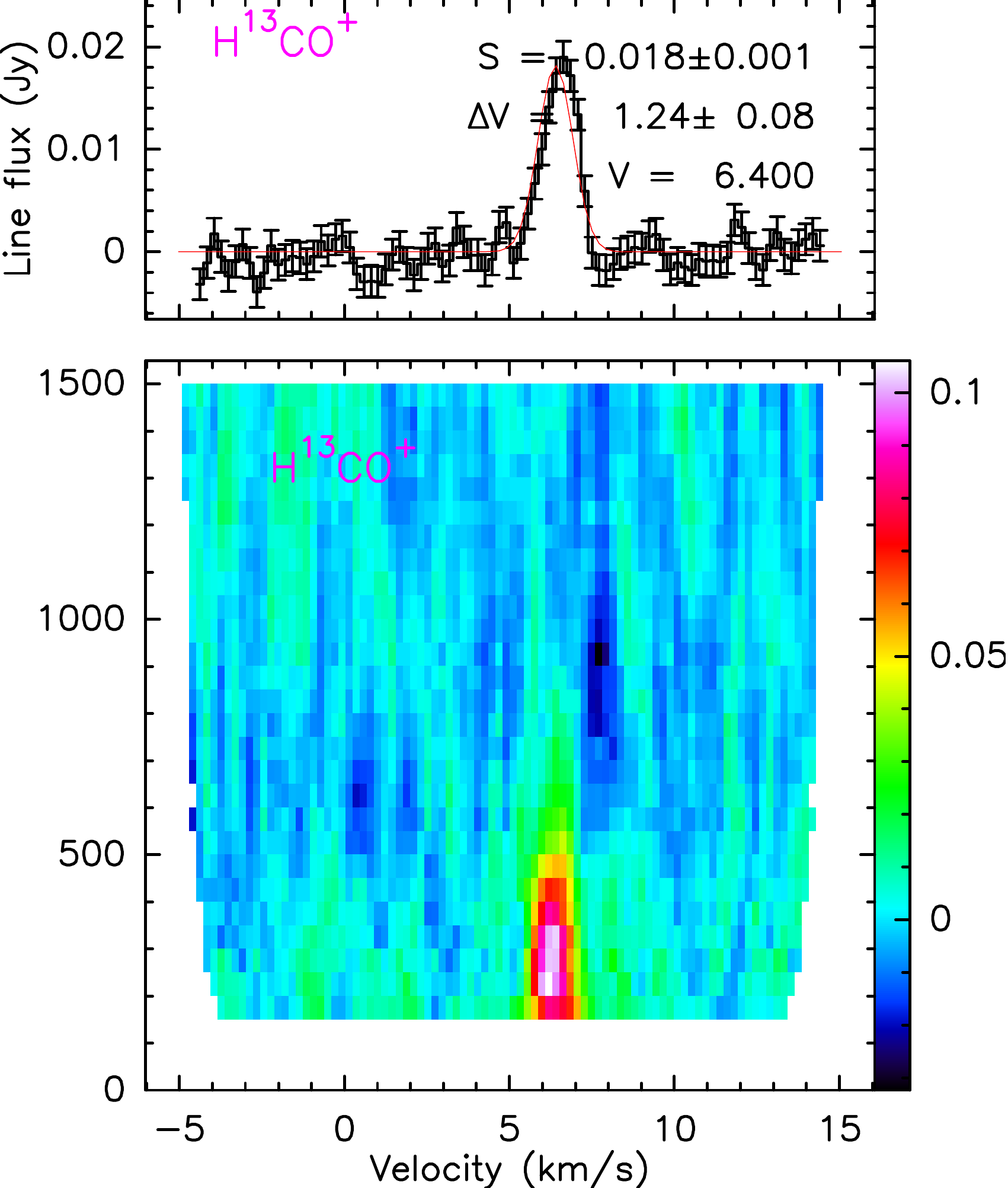}
 \includegraphics[width=0.3\linewidth]{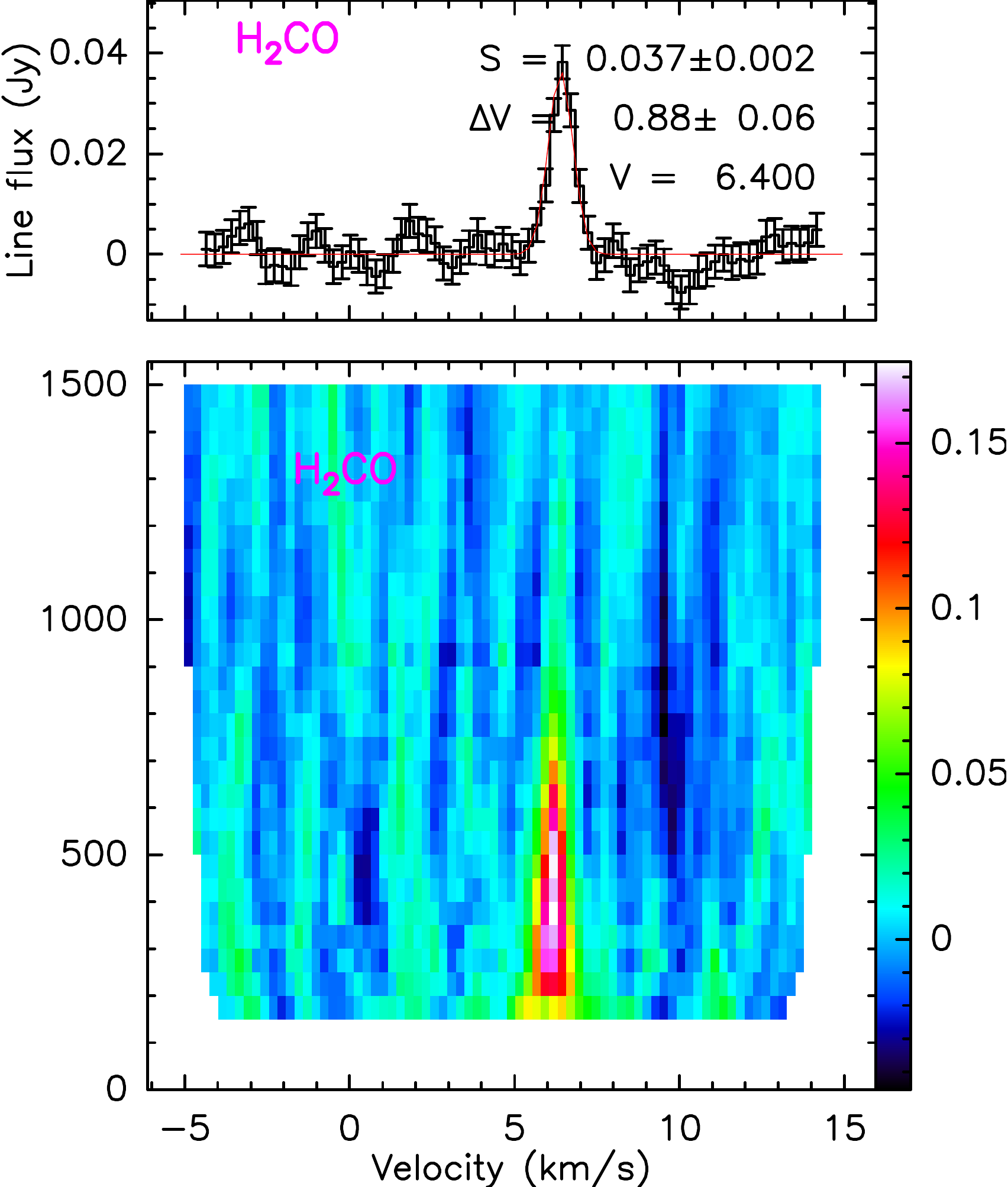}
 \includegraphics[width=0.31\linewidth]{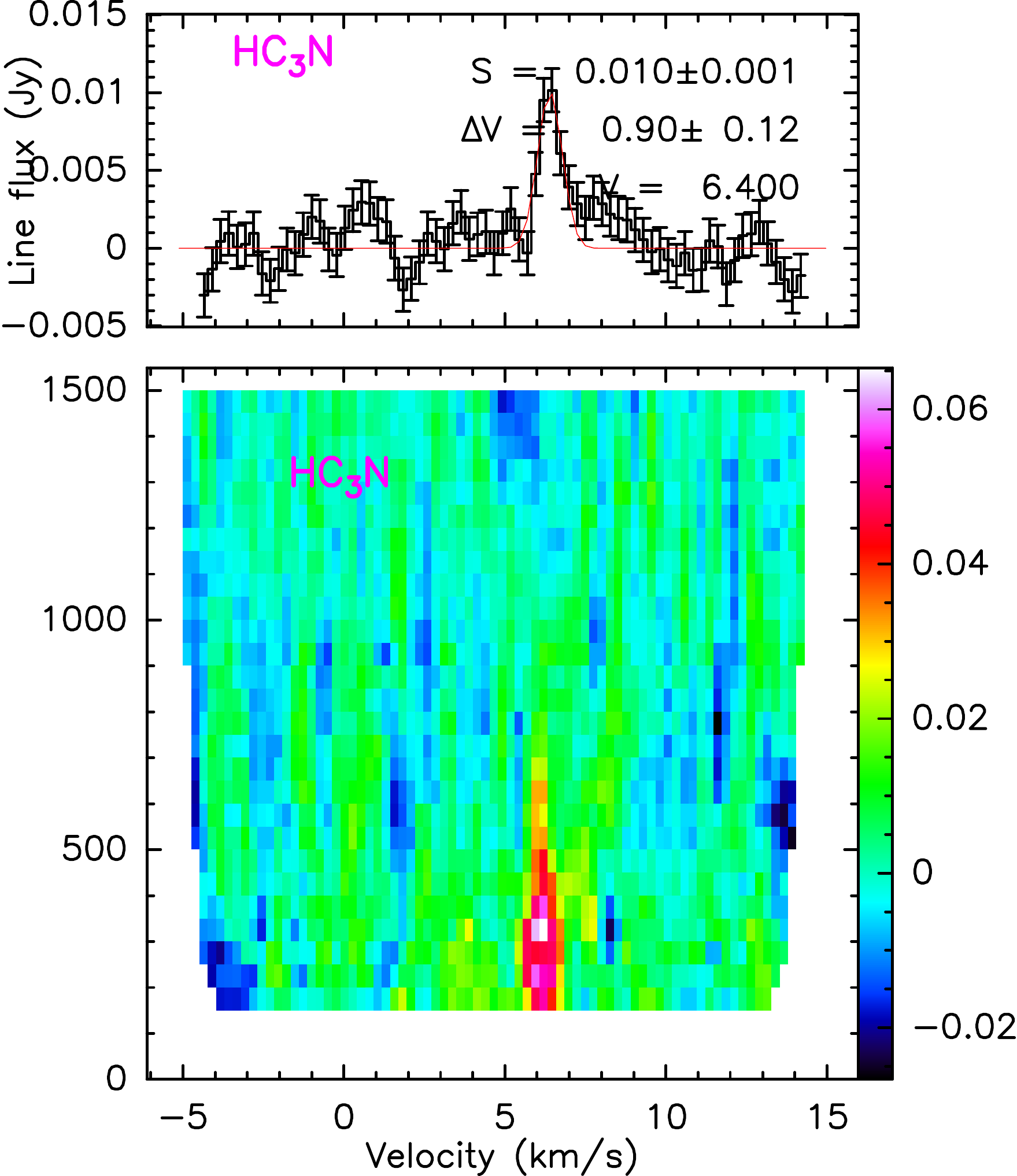}
 \caption{\textit{Top:} Integrated spectra of CN, HCO$^+$, HCN, N$_2$H$^+$, HNC, CCH, H$^{13}$CO$^{+}$, H$_2$CO, and HC$_3$N (with Gaussian fits, including hyperfine structure when needed, in red). The values indicate the fit results. \textit{Bottom:} Corresponding radial-velocity diagram obtained after making Keplerian velocity correction and line stacking.
 } \label{fig:rv}
 \end{figure*}
\end{appendix}

%
%

\end{document}